\documentclass{aa}  
\usepackage{natbib}
\bibpunct{(}{)}{;}{a}{}{,}
\usepackage{graphicx, subcaption}
\usepackage{placeins}
\usepackage{txfonts}
\usepackage{ulem}
\usepackage[colorlinks=true,     linkcolor=blue, citecolor=blue, filecolor=blue, urlcolor=blue]{hyperref}
\usepackage{placeins}

\newcommand{\e}[1]{\,$\times\,10^{#1}$}
\newcommand{\Fxuv}[0]{$F_{\text{XUV}}$}
\newcommand{\uFxuv}[0]{erg\,cm$^{-2}$\,s$^{-1}$}
\newcommand{\vFxuv}[1]{\Fxuv$=#1$ \uFxuv}
\newcommand{\HH}[0]{$\text{H}_2$\;}

\begin{document} 

        \title{Radio emission from close-in exoplanets: Can we extend the radio-magnetic scaling law to the sub-Alfvénic stellar wind regime?}
   \titlerunning{The radio-magnetic scaling law in the sub-Alfvénic regime}
    \authorrunning{Elekes \& Vidotto}
        
        \author{Filip Elekes
                \and
                Aline A. Vidotto
        }
        
        \institute{Leiden Observatory, Leiden University, Einsteinweg 55, NL-2333 CC Leiden\\ 
                \email{elekes@strw.leidenuniv.nl}        
        }
        
        \date{Received 7 April 2026 / Accepted 6 July 2026}

        \abstract
    {Observations of exoplanetary radio auroras can directly probe planetary magnetic fields and be used to infer properties of the local magnetic star--planet interaction. However, the search for exoplanetary radio auroras has been ongoing for decades without indisputable detections despite numerous favorable predictions based on extrapolations of the radio-magnetic scaling law (RMSL).}
    {The RMSL is based on Solar System planets that lie in the super-Alfvénic solar wind regime, and it is unclear whether the relation holds for close-in exoplanets residing in sub-Alfvénic and strongly magnetized winds. We aim to test whether the relation can be extended to sub-Alfvénic stellar winds.}
        {We employed 3D magnetohydrodynamic simulations of the magnetosphere of a Jupiter-like planet at various distances from the Sun to study properties of the expected radio emission, taking the atmospheric photoionization into consideration and using a solar wind model with a nominal and an enhanced magnetic field.}
        {Our radio predictions match the RMSL in the super-Alfvénic solar wind regime. However, we find that in the sub-Alfvénic solar wind regime, RMSL predictions overestimate the planetary radio output by one order of magnitude. This discrepancy is significantly higher in the case of a stronger magnetized wind, which is caused by a strong decrease in wind-magnetosphere energy transfer efficiency. Thus, planetary radio emission may fall below the sensitivity limit of current radio telescopes, which could partially explain the lack of detections. We expect the overestimation by the RMSL to increase even further with more magnetic cool stars. Furthermore, due to atmospheric photoionization and the resulting high ionospheric electron densities, favorable conditions for the generation and escape of electron-cyclotron maser instability (ECMI)-driven radio emission are constrained to larger orbital distances ($\gtrsim 0.1\,$au) and to the planetary nightside in the solar wind case. This further decreases our predicted radio powers by up to one order of magnitude. In more strongly magnetized winds, enhanced open planetary magnetic flux and reconnection-driven outflows cause magnetospheric electron depletion, resulting in improved ECMI conditions and effectively eliminating the mitigating effect of strong photoionization on ECMI emission.
    Via calibration with the RMSL, we find the combined electron acceleration and ECMI efficiency in the super-Alfvénic solar wind to be $\approx10^{-3}$. We argue that the efficiency is unlikely to increase up to $\approx10^{-1}$ in sub-Alfvénic regimes, which would be needed to overcome the predicted discrepancy between RMSL and our predictions.
    }
    {}
        
        \maketitle
    \nolinenumbers

        \section{Introduction}
    Star--planet magnetic interactions (SPMIs) have the potential to release large energy fluxes, with possible far-reaching consequences for the evolution of planets, atmospheres, and their magnetospheres. Strong SPMIs can lead to observable signatures that enable the inference of local conditions at the stars, their winds, or the planets. Compact exoplanetary systems represent an exceptional laboratory for probing such SPMIs due to the high energy densities involved in the stellar outflows (e.g., winds and eruptive phenomena) to which close-in exoplanets are exposed \citep{Vidotto2015,Elekes2025}. To study and characterize such SPMIs, knowledge about stellar wind properties (e.g., plasma density, magnetic field, and velocity) and planetary characteristics (e.g., magnetic field and atmosphere properties) are necessary. While stellar winds can be partially constrained using stellar magnetic field maps \citep[e.g.,][]{Vidotto2009,Evensberget2025} obtained from spectropolarimetric observations and the Zeeman-Doppler imaging method \citep[e.g.,][]{Semel1989,Donati1997}, knowledge about the planetary magnetic fields is still largely unconstrained. Planets represent an obstacle in the ionized stellar wind flow, and the amount of energy involved in such SPMIs is largely determined by the size of the obstacle and its ability to effectively convert the incoming energy fluxes. Planetary magnetic fields determine the size of the obstacle (i.e. the magnetosphere) and provide generator mechanisms capable of converting the incoming energy fluxes into other forms of energy \citep{Akasofu1981,Elekes2025}. From the Solar System planets, we know that part of this energy is released in auroral emissions.
    
    Observations of planetary auroral radio emission are particularly important as they can be used to directly infer the source region's magnetic field strength due to the emission frequency being dictated by the local electron gyrofrequency \citep{Hess2011,Kavanagh2024,Callingham2024}. The generation of such emissions is generally attributed to the electron cyclotron maser instability (ECMI) mechanism, whose efficiency is sensitive to local plasma conditions (e.g., the electron density and magnetic field strength; \citealt{Winglee1986,Zarka1998,Treumann2006}). Planetary auroral radio emissions have been observed from all magnetized planets in the Solar System. For example, Jupiter's magnetic field was constrained by radio observations long before it was measured in situ by space probes \citep{Franklin1958}. Consequently, great efforts have been made over decades to observe such exoplanetary radio emissions, but so far without unambiguous success \citep[e.g.,][]{Yantis1977,Winglee1986b,Bastian2000,Lazio2007,Hallinan2013,Zarka2015,Murphy2015,Lynch2017,Bastian2018,deGasperin2020}.
    The most promising observation was reported recently \citep[$\tau$\,Boo\,b;][]{Turner2021}. However, this measurement has not yet been confirmed by follow-up detections \citep{Turner2023,Turner2024,Cordun2025}.

    Naturally, the question arises as to why the search for exoplanetary radio emissions is proving so difficult. There are several possible reasons: 
    The radio emission is too weak, or the sensitivity of current radio telescopes is too low \citep{Bastian2000,Vidotto2015,Lynch2017,Elekes2023,Evensberget2025}. The bandwidth of possible radio emissions is large, while the bandwidth of individual telescopes is relatively small. Hence, it is possible that emissions have not been detected due to a mismatch in frequency bands \citep{Bastian2000}.
    The Earth's ionosphere is impervious to radio waves with frequencies below 10 MHz \citep[the ionospheric cutoff;][]{Wilson2013}. 
    A similar situation may exist at the exoplanets since high plasma densities in the magnetosphere caused, for example, by inflated atmospheres or strong stellar X-ray and ultraviolet (XUV) radiation and associated atmospheric escape can hinder auroral radio emission to escape \citep{Weber2017,Weber2018,Daley-Yates2018}. The ECMI mechanism is also known to operate at reduced efficiency if the local gyro frequency is too close to or lower than the local plasma frequency \citep{Treumann2006}. Thus, the primary targets of the search for exo-radio auroras, i.e., close-in giant planets \citep{Griessmeier2007b,Jardine2008}, may prove to be worse radio emitters than previously anticipated \citep{Griessmeier2017}. Geometric causes may also be a factor: 
    auroral ECMI emission is beamed along hollow cones centered at high-latitude polar magnetic field lines located within so-called auroral ovals \citep[e.g.,][]{Zarka1998,Farrell1999,Zarka2007}. Because of this as well as the rotation and orbital movement of the planet, it may be difficult to align the radio beam with our line of sight \citep{Bastian2000,Kavanagh2023}. The magnetic topology of the star--planet interaction also plays a major role in dictating the amount of converted energy \citep{Vidotto2012,Strugarek2015,See2015}. The energy converted in the magnetospheres can vary by orders of magnitude depending on the angle between the stellar and planetary magnetic fields and consequently the amount of magnetic reconnection happening between the wind and the planetary magnetic field \citep{Strugarek2015,Elekes2023}. Magnetically active stars in particular create difficult conditions for such observations, as magnetic polarity reversals \citep[e.g., at $\tau$ Boötis b;][]{Jeffers2018,Turner2021} or transient changes in space weather and stellar outflows \citep[winds, prominence eruptions, and coronal mass ejections; e.g.,][]{Callingham2024} can significantly influence the radiated power and, consequently, result in the beamed planetary emissions having a periodic and erratic on-off nature  \citep{Jardine2008,Vidotto2012,See2015}. Thus, successful follow-up observations may prove exceptionally difficult because of the varying conditions at the exoplanets.
    
    Radio observations in the Solar System have shown that the auroral radio powers of all magnetized planets and Ganymede indicate a linear dependence on the Poynting flux of the ambient plasma flow (i.e., the solar wind or magnetospheric plasma; \citealt{Farrell1999,Zarka2007,Zarka2018}).
    This linear correlation has often been used to predict exoplanetary radio emissions \citep[e.g.,][]{Farrell2003,Lazio2004,Stevens2005,Burkhart2017}. Extrapolation of this so-called radio-magnetic scaling law (RMSL; also known as the radiometric Bode law) suggests that especially strongly magnetized gaseous planets close to their host stars should be strong radio emitters, with emitted powers exceeding those of the Solar System planets by orders of magnitude. Consequently, the search for exo-radio auroras has been largely focused on close-in gas giants \citep{Griessmeier2017,Lynch2018,Weber2018}. However, with the exception of the Jovian moons, the RMSL is based on planets situated in super-Alfvénic solar wind conditions with constant energy partition. Although sub-Alfvénic conditions prevail in Jupiter’s magnetosphere, our understanding of these conditions cannot necessarily be applied to sub-Alfvénic stellar wind conditions since the Mach numbers, plasma conditions, and general magnetic topology can vary significantly, which would have a great effect on the SPMI \citep{Paul2026}.
    It is assumed that a large number of close-in exoplanets are exposed to sub-Alfvénic winds \citep{Saur2013,Zhilkin2019}, and it is not clear whether an extrapolation of the RMSL from super- to sub-Alfvénic conditions is valid. Doubts as to whether the RMSL overestimates radio power have been expressed in the literature \citep{Nichols2016,Turnpenney2020,Cordun2025}.

    We explored the energetics of exoplanetary radio emissions caused by SPMIs in super- and sub-Alfvénic stellar wind regimes by means of 3D magnetohydrodynamic (MHD) simulations. By studying the relevant energy fluxes, we address the question of whether the RMSL can be extrapolated to a sub-Alfvénic wind regime. We took the Solar System as an example and studied the energetics of SPMIs as we pushed a Jupiter-like planet closer to the Sun, thus increasing the incident stellar wind power and the stellar XUV flux, which in turn increases the magnetospheric electron population via atmospheric photoionization, possibly hindering the generation and escape of radio waves \citep{Treumann2006,Weber2017}. The paper is structured as follows: In Sect. \ref{Sect:Simulation} we describe the stellar wind and planetary MHD models. In Sect. \ref{Sect:Radio_RMSL_definition} we describe the RMSL and our method for calculating the radio powers. We present our prediction of the RMSL extrapolation toward the sub-Alfvénic solar wind regime in Sect. \ref{Sect:Radio_magnetic_scaling_law}, which is followed by a discussion on the energetics of the SPMIs and the characteristics of auroral radio source regions in Sect. \ref{Sect:Energetics_and_source_regions} and our conclusions in Sect. \ref{Sect:Conclusions}.

        \section{Numerical simulation} \label{Sect:Simulation}
    
    \begin{table*}[!htb]
    
    \begin{minipage}{.6\linewidth}

      \caption{Physical simulation parameters.}      
                \label{Tab:Parameters_fixed}      
                \centering                                   
                \begin{tabular}{c c c c}          
                        \hline\hline                       
                        Quantity &  Symbol & Value & Unit\\    
                        \hline\hline                              
                        Stellar model & & &  \\ \hline
            Stellar radius & $R_\star$ & 1& $R_\odot$   \\
            Stellar mass & $M_\star$ & 1 & $ M_\odot$   \\
                        Rotation rate & $\Omega_\star$ & $2.67$\e{-6} & rad/s   \\
                        Radial magnetic field (B1) & $B_{\star}$ & 2.8 & G   \\
            Radial magnetic field (B10) & $B_{\star}$ & 28 & G   \\
                        Base coronal density & $n_{\star}$ & $1.25$\e{8} & cm$^{-3}$   \\
            Base coronal temperature & $T_{\star}$ & 1.535 & MK   \\
                        XUV flux at $0.03$\,au & $F_{\rm XUV}$ & 2200 & erg\,cm$^{-2}$\,s$^{-1}$   \\  
                        \hline\hline         
            Planet model & & &   \\ \hline
            Planet radius       & $R_p$ & 69911 & km   \\      
                        Atmosphere base density & $n_0$ & $10^6$ & H$_2$ cm$^{-3}$  \\
            Atmosphere scale height & H & 0.06 & $R_p$  \\
            H$_2$-H$^+$ collisional cross section & $\sigma_\text{coll}$ & $2$\e{-15} & cm$^2$  \\
            Photo-ion. cross section & $\sigma_\text{ph}$ & $10^{-17}$ & cm$^2$  \\
            Magnetic field strength (Eq.) & $B_p$ & 1 & G \\
                        \hline
            
                \end{tabular}
        
    \end{minipage}
    \begin{minipage}{.4\linewidth}

      \caption{Varied planetary parameters.}      
                \label{Tab:Parameters_variable}      
                \centering                                   
                \begin{tabular}{c c c c}          
                        \hline\hline                       
                        a  &  \Fxuv & $T_e^{\star}$ & $\alpha_{\rm rec}^{\dagger}$ \\    
              $\left[\text{au}\right]$     &  [\uFxuv] & [K] & [cm$^2$\,s$^{-1}$] \\  
                        \hline                            
                        
            0.03 & 2200 & 7500 & 5.098\e{-13} \\
            0.04 & 1238 & 7500 & 5.098\e{-13} \\
            0.05 & 792.0 & 7500 & 5.098\e{-13} \\
            0.07 & 404.1 & 7200 & 5.233\e{-13} \\
            0.1 & 198.0 & 7200 & 5.233\e{-13} \\
            0.2 & 49.50 & 1700 & 1.318\e{-12} \\
            0.4 & 12.37 & 1500 & 1.428\e{-12} \\
            0.7 & 4.041 & 1400 & 1.492\e{-12} \\
            1.0 & 1.980 & 1200 & 1.647\e{-12} \\
            2.0 & 0.495 & 1200 & 1.647\e{-12} \\
            
                        \hline
            
                \end{tabular}
        \tablefoot{\\
        $\star$: Estimated from \citet{Koskinen2010}.\\
        $^\dagger$: Calculated with $\alpha_{rec}=4\times10^{-12} \left(\frac{300\,\text{K}}{T_e}\right)^{0.64}$ \citep{Storey1995}.
        }

    \end{minipage} 
    
    \end{table*}
    
    To study stellar wind-planet interactions and associated planetary auroral radio emissions, we need descriptions of the stellar wind and the planetary magnetosphere interacting with the wind. We describe our stellar wind in Sect. \ref{Sect:Stellar_wind_model} and planetary models in Sect. \ref{Sect:Planetary_MHD_model}, followed by a brief description of the simulated stellar winds (Sect. \ref{Sect:Structure_energetics_wind}) and planetary magnetosphere (Sect. \ref{Sect:Structure_SPI}).

    \subsection{Stellar wind}\label{Sect:Stellar_wind}
    Our study of the stellar wind-planet interaction comprises two different stellar wind models. Our basic stellar wind model is a representation of the nominal solar wind. 
    Many target exoplanets arguably orbit more strongly magnetized stars. Hence, we additionally performed stellar wind simulations with a stronger stellar magnetic field.

    \subsubsection{Stellar wind model}\label{Sect:Stellar_wind_model}

    We employed the Versatile Advection Code \citep[VAC 4.5;][]{Keppens1999} to solve the 1.5D MHD equations for a thermally driven magneto-rotator wind \citep{Weber1967} in spherical coordinates until a steady state solution was achieved. We solved for the density, the temperature, and the radial and angular components of the magnetic field and velocity that depend only on the radial distance. Our model is similar to \citet{Johnstone2015a}, except that we kept a fixed value of $1.05$ for the polytropic index. The boundary conditions for the wind simulations are based on solar properties, whereby the free parameters, namely the coronal base density $n_\star$, temperature $T_\star$ and radial magnetic field strength $B_\star$ (see Table \ref{Tab:Parameters_fixed}), are adjusted in such a way that the stellar wind solution results in the solar mass loss rate of $\dot{M}_\star \approx 3\times 10^{-14}$\,M$_\odot$ and terminal wind velocity between $400$ and $500$ km/s beyond $1$\,au. The chosen coronal radial magnetic field strength of $B_\star=2.8$\,G lies within the range of observed solar values \citep{Johnstone2015b}. The Alfvén radius is at $\approx 0.11$\,au $\approx 21\, R_\odot$, which is placed near the upper limit of observed solar wind Alfvén radii \citep{Kasper2021,Cranmer2023,Badman2025}. We refer to this solar wind model as the B1 model. The physical simulation parameters of the stellar wind simulations can be found in Table \ref{Tab:Parameters_fixed}.
    In addition to the solar wind model, we also considered a stellar wind model with a base magnetic field strength ten times that of the solar model, $B_\star^{\rm B10} = 28$\,G, hereinafter referred to as the B10 model. By increasing the magnetic field strength, the Alfvén radius increased to $\approx 1.5$\,au. The plasma density, pressure, and velocity were only negligibly affected. This allowed us to study the effect of a stronger magnetized stellar wind on possible auroral radio emission in as isolated a manner as possible, without significantly influencing the wind's kinetic, thermal, and structural properties.
    We note that, in reality, more magnetically active stars should have enhanced coronal magnetic fields that are accompanied by enhancements in mass-loss rates and temperature, which would affect not only the magnetic field of the wind but also its density and velocity. In this study we aimed to test the effect of the wind's magnetic-to-kinetic energy ratio on planetary auroral energetics in an isolated manner, which seems to be the root cause for the validity of the RMSL in the solar wind.

    The plasma parameters of both stellar wind models as a function of distance from the star are shown in Appendix \ref{Sect:Appendix:Structure_wind}.

    \begin{figure}
        \centering
        \includegraphics[width=1\linewidth]{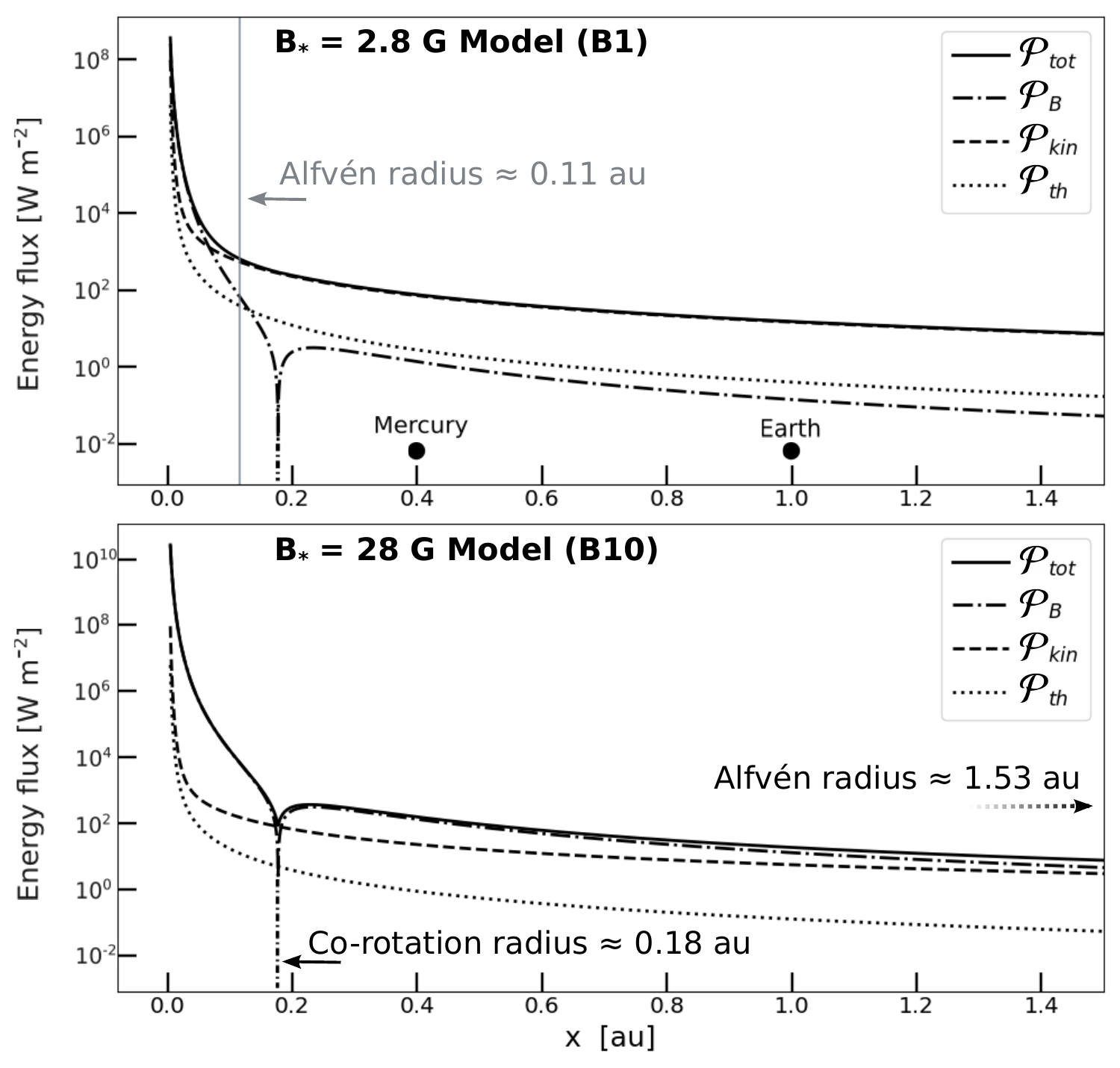}
        \caption{Stellar wind energy fluxes (W/m$^2$) in our basic ($B_\odot=2.8$ G) and up-scaled magnetic field model ($B_\odot=28$ G) at various orbital distances (in au). We show the stellar wind Poynting flux ($\mathcal{P}_{\rm B}$) and the kinetic ($\mathcal{P}_{\rm kin}$), thermal ($\mathcal{P}_{\rm th}$), and total ($\mathcal{P}_{\rm tot}$) energy flux in the reference frame of the planets orbiting at a distance of x from the star (Eqs. \ref{Eq:Poynting_flux_wind}-\ref{Eq:Thermal_energy_flux}).} 
        \label{Fig:Wind_energy}
    \end{figure}

    \subsubsection{Structure and energetics of the stellar wind}\label{Sect:Structure_energetics_wind}
    
    In Fig. \ref{Fig:Wind_energy} we show the energy fluxes of the modeled stellar winds in the frame of a planet in a circular orbit as a function of orbital distance from the star. The Poynting flux $\mathcal{P}_{\rm B}$, kinetic $\mathcal{P}_{\rm kin}$ and thermal $\mathcal{P}_{\rm th}$ energy fluxes are given by the following equations:
    \begin{align} 
        \mathcal{P}_{\rm B}     &= \frac{B_{\rm sw}^2}{\mu_0} v_\perp \label{Eq:Poynting_flux_wind} \\
        \mathcal{P}_{\rm kin} &= \frac{1}{2} \rho_{\rm sw}\, v_{\rm 0}^3 \label{Eq:Kinetic_energy_flux} \\
        \mathcal{P}_{\rm th}  &= n_{\rm sw}\,k_B T_{\rm sw} v_{\rm 0} \label{Eq:Thermal_energy_flux} \; ,
    \end{align}
    where $\rho_{\rm sw}(x)$ is the stellar wind plasma density (and  $n_{\rm sw}$ the number density), $B_{\rm sw}(x)$ the magnetic field, $v_{\rm 0}(x) $ the relative velocity, and $T_{\rm sw}(x)$ the temperature. The relative velocity in Eqs. \ref{Eq:Poynting_flux_wind}-\ref{Eq:Thermal_energy_flux} is defined as $\vec{v}_{\rm 0}=\vec{v} - \vec{v}_{\rm orb}$, where $\vec{v}$ is the wind and $\vec{v}_{\rm orb}$ the planet's Keplerian velocity at distance $x$. The relative velocity perpendicular to the wind magnetic field is denoted by $v_{\perp}$.

    Our B1 stellar wind model reproduces the solar wind energy partition between kinetic energy and Poynting flux. Thus, the kinetic energy flux is on the order of 100 times higher than the wind's Poynting flux beyond $1$\,au and further out \citep{Zarka2007}. The stellar wind kinetic energy and Poynting flux decay with $1/r$ within the asymptotic regime beyond $1$\,au. The wind's Parker spiral was reproduced as well. At $1$\,au, the stellar wind magnetic field is at approximately 40 degrees with respect to the star-planet line. The wind speed beyond $1$\,au is between $400$ and $500$\,km/s and the magnetic field strength at $1$\,au is approximately $10^{-4}$\,G, consistent with solar wind values \citep{Bagenal2013}. 
    At the corotation radius near $0.18$\,au, the wind's magnetic field is aligned with the flow velocity, resulting in a minimum Poynting flux (see Eq. \ref{Eq:Poynting_flux_wind}).
    
    The B10 model conserves the structure of the B1 solar wind model. However, the stellar wind Poynting flux is consistently enhanced by two orders of magnitude due to the increase in stellar magnetic field strength by a factor of $10$ (since $\mathcal{P}_{\rm B} \propto B_{\rm sw}^2$). The Alfvén radius is shifted to $1.53$\,au, which lies outside of our parameter space for the subsequent simulations.

        \subsection{Planetary magnetosphere}\label{Sect:Planetary_magnetosphere}
    Our planet-centered SPMI model encompasses the interaction of the planetary magnetosphere with the surrounding steady-state stellar wind. We used the simulated stellar wind properties (see \ref{Fig:Stellar_wind_structure}) as constant boundary conditions for the SPMI simulations. In the following we describe the MHD model, the parameterizations of physical processes governing plasma production and loss, and show the structure of the modeled magnetospheres.

    \subsubsection{Planetary MHD model}\label{Sect:Planetary_MHD_model}

    We modeled the interaction of a planetary magnetosphere with a steady-state wind using a 3D ideal\footnote{With ideal MHD, plasma resistivity is solely driven by numerical diffusion imposed by the numerical scheme. We refer to \citet{Elekes2023}, where we tested the effect of physical resistivity on the simulated quantities in a similar simulation setup and found it to be negligible.} 
    MHD model. We used the PLUTO code \citep[v. 4.4;][]{Mignone2007} to numerically solve the following set of ideal MHD equations in spherical coordinates,
        \begin{eqnarray}
                \frac{\partial \rho}{\partial t} + \nabla \cdot\left[\rho \vec{v}\right] &=& Pm_n - L m_p \label{Eq:Continuity_eq} \\
                \frac{\partial \rho \vec{v}}{\partial t} + \nabla \cdot \left[ \rho \vec{v} \vec{v}  + p- \vec{B} \vec{B} + \frac{1}{2}  B^{2} \right] &=& -(L m_p + \nu_n \rho)\vec{v} \label{Eq:Momentum_eq} \\
                \frac{\partial E_t}{\partial t} + \nabla \cdot \left[(E_t + p_t)\vec{v} - \vec{B}(\vec{v}\cdot\vec{B}) \right] &=& -\frac{1}{2} (L m_p + \nu_n \rho)v^{2} \nonumber 
            \\ & & -\frac{3}{2} (L m_p + \nu_n \rho) \frac{p}{\rho} \nonumber 
            \\ & & +\frac{3}{2} (P m_n + \nu_n \rho) \frac{k_B T_n}{m_n} \label{Eq:Energy_eq}\\
                \frac{\partial \vec{B}}{\partial t}- \nabla \times \left[\vec{v} \times \vec{B}\right] &=&0 \label{Eq:Induction_eq},
        \end{eqnarray}
    where $\rho$ is the plasma density, $\vec{v}$ is the velocity, $\vec{B}$ is the magnetic field and $p$ is the thermal pressure.  
    The total energy density is $E_t = \rho e + \rho v^2 /2+ B^2 / 2\mu_0$, where $e$ is the specific internal energy and $p_t$ the total pressure (e.g., the sum of magnetic and thermal pressure). We used the adiabatic equation of state, $p = \rho e (\gamma - 1)$, to close the system, where $\gamma=5/3$ is the ratio of specific heats.
    In the right hand side of Eqs. \ref{Eq:Continuity_eq}-\ref{Eq:Energy_eq} we include source terms associated with elastic ion-neutral collisions caused by the presence of a neutral atmosphere to account for momentum transfer, internal energy transfer, and charge exchange.
    There, $P$ and $L$ denote plasma production and loss terms, respectively. The quantities $m_n$ and $m_p$ are the masses of neutral and plasma particles, respectively. The physical parametrization of $P$ and $L$ will be described in the following paragraphs.

    \paragraph{Atmosphere and magnetic field}
    The parameters of the planetary model discussed here can be found in Tables \ref{Tab:Parameters_fixed} and \ref{Tab:Parameters_variable}. We considered a radially symmetric neutral atmosphere consisting of molecular hydrogen H$_2$ with the density following a barometric law,
    \begin{equation}\label{Eq:Atmosphere_density_barometric}
                n_{n}(r) = n_{n,0} \exp\left(\frac{R_p - r}{H}\right) \;,
        \end{equation}
    where $n_n$ is the neutral number density, $r$ is the distance from the planet's center and $H=0.06\,R_p$ is the scale height, where $R_p=69911$\,km is the planet's radius (i.e., one Jupiter radius). The scale height was chosen to sufficiently resolve the atmosphere with our numerical grid. At the inner simulation boundary we assumed a neutral number density of $n_{n,0}=1$\e{6} cm$^{-3}$, which roughly corresponds to an upper atmosphere with pressure in the millibar range. Collisions between the static atmosphere (i.e., it does not evolve during the simulation) and the plasma are modeled using a collision frequency $\nu_c$ on the order of 1 s$^{-1}$, so that $\nu_n \approx \sigma_c \, n_{n}(r)\, \Bar{v}$, with a characteristic plasma velocity $\Bar{v}$ and the ion-neutral collisional cross section $\sigma_c=2$\e{-15} cm$^{2}$ \citep[e.g.,][]{Johnstone2018}. Ion-neutral collisions pose a sink of kinetic energy that is included in Eqs. \ref{Eq:Momentum_eq} and \ref{Eq:Energy_eq}. We also accounted for charge-exchange between plasma and neutral particles and associated exchange of internal energy between the colliding particles (Eq. \ref{Eq:Energy_eq}).

    \begin{figure*}
        \centering
        \includegraphics[width=1\linewidth]{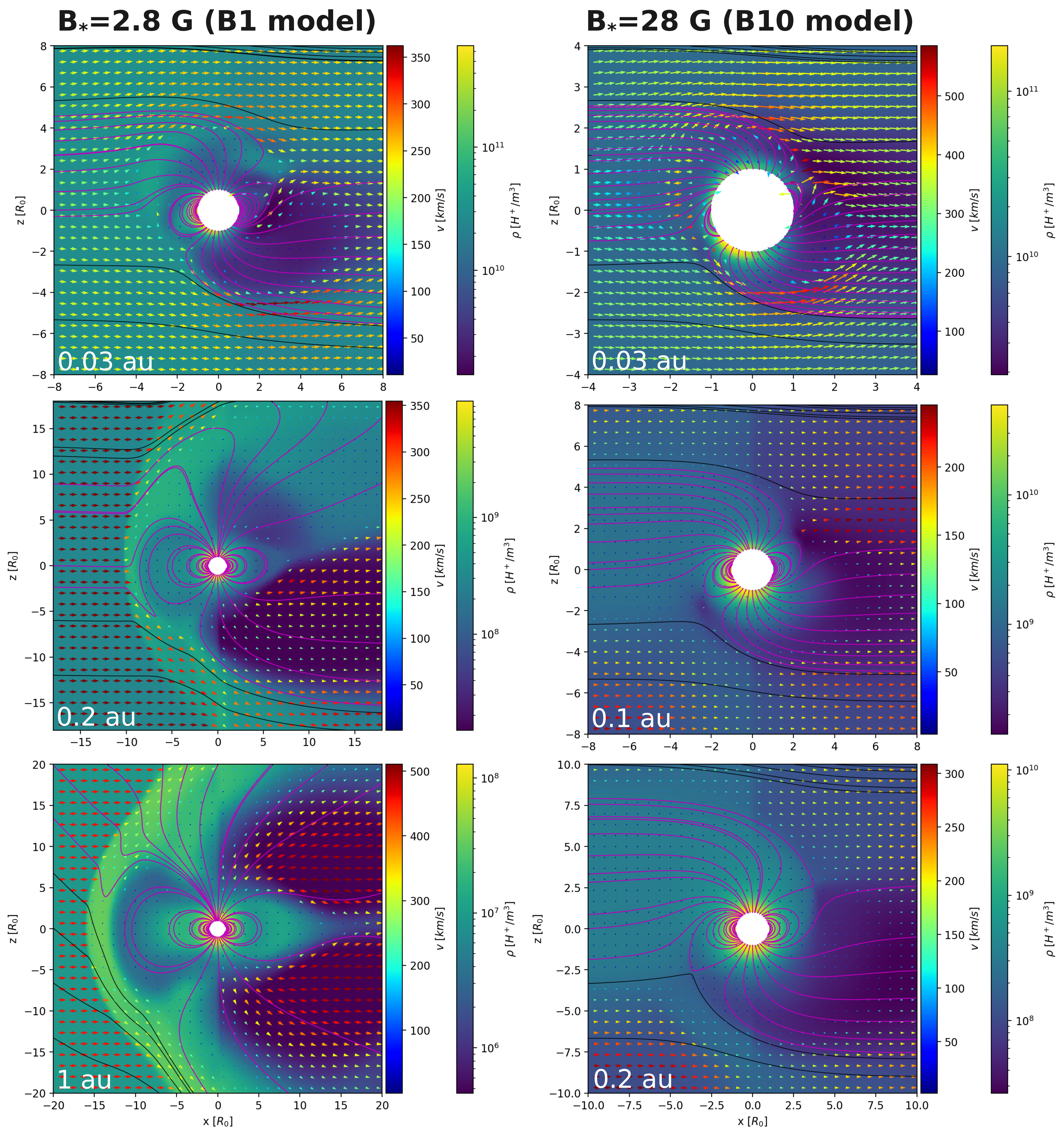}
        \caption{XZ plane cross sections of the simulation domain for planets at various orbital distances. Arrows indicate velocity vectors and their magnitudes (left colorbar). Contours show plasma density (right colorbar). Solid lines show magnetic field line projections. We show the results for the basic (B1; \textit{left}) and enhanced magnetic field model (B10; \textit{right}). The panels show different zoom levels, as the magnetospheres have different sizes.}
        \label{Fig:Magnetosphere_structure}
    \end{figure*}
    Photoionization through XUV radiation is the main driver of atmospheric escape in hydrogen-rich atmospheres \citep[e.g.,][]{Lammer2003} whose rates typically exceed those of other ionization processes such as electron impact ionization \citep[e.g.,][]{Voronov1997}.
    Inspired by this, plasma production in our model is solely driven by photoionization of the neutral \HH atmosphere. The production term ($P$) in Eqs. \ref{Eq:Continuity_eq}-\ref{Eq:Energy_eq} is defined as
    \begin{equation}
        P = \sigma_{\rm ph}\, n_n(r)\, F_{\rm XUV}\, e^{-\tau}\;,
    \end{equation}
    where $\sigma_{\rm ph}$ is the photo-absorption cross section of the neutral particles and $F_{\rm XUV}$ is the stellar incident XUV flux. The optical depth $\tau$ is integrated along a line starting from the XUV source at the outer, substellar simulation boundary at $\vec{r}_0$, to the respective position within the neutral atmosphere at $\vec{r}$, 
    \begin{equation}\label{Eq:OpticalDepth}
        \tau= \int_{\vec{r}_0}^{\vec{r}} n_n(r) \, \sigma_{\rm ph}\, \text{d} \vec{r}\; .
    \end{equation}
    In Appendix \ref{Sect:Appendix:CoordinateSystem} we provide a schematic illustrating to illustrate the geometry of the coordinate system and optical depth integration.
    We assumed an XUV flux of \vFxuv{2200} at a distance of $0.03$\,au from the star. This value is placed within the lower range of typical XUV fluxes of Sun-like stars \citep{SanzForcada2011} and slightly above the respective XUV flux of the Sun \citep{Ribas2005}. We furthermore assumed that the entire XUV energy is concentrated at $16.7$\,eV to maximize the \HH photoionization cross section, which accordingly amounts to $\sigma_{\rm ph}=1$\e{-17}\,cm$^{2}$, slightly above the ionization threshold at $15.44$\,eV \citep[][]{Heays2017}. The estimated maximum ionization rate at $1$\,au with an atmosphere base density of $n_n = 10^6$\,\HH\,cm$^{-3}$ is approximately $9$\e{-7}\,s$^{-1}$, which was required for the simulations to be stable and sufficiently fast.
    
    We considered plasma loss due to dissociative recombination of ions with free electrons. The recombination rate of hydrogen ions, $\alpha_{\rm rec}$ (Table \ref{Tab:Parameters_variable}), was obtained from \citet{Storey1995} by assuming electron temperatures $T_e$ for planets at different orbital positions in the range $1200$ - $7500$\,K \citep[][]{Koskinen2010}. The recombination rate depends only weakly on $T_e$ and thus the uncertainties of our assumed values have no significant effect on the results of this work. The associated loss terms in Eqs. \ref{Eq:Continuity_eq}-\ref{Eq:Energy_eq} are defined as
    \begin{equation}\label{Eq:Recombination}
        L=\alpha_{\rm rec} n\left(\vec{r})(n(\vec{r})-n_{\rm sw}\right)\;
    \end{equation}
    and also account for associated cooling and kinetic energy loss of the plasma (Eq. \ref{Eq:Energy_eq}).
    Recombination is switched off if the plasma density falls below the background stellar wind density $n_{\rm sw}$.

    We assumed a dipolar planetary magnetic field with its axis perpendicular to the orbital plane. The equatorial field strength was set to $1\,G$ for all models.

    \paragraph{Numerical model}
    Equations \ref{Eq:Continuity_eq}-\ref{Eq:Induction_eq} were integrated using the Harten-Lax-Van Leer Riemann solver with the minmod flux limiter. Time-stepping was achieved with the second order Runge-Kutta method. To ensure the $\nabla\,\cdot B=0$ condition, the extended mixed hyperbolic-parabolic divergence cleaning technique was utilized \citep{Dedner2002,Mignone2010}.
    The numerical schemes used in this study pose a particularly diffuse setup. The diffusivity introduced by the scheme directly affects magnetic reconnection rates and thus likely affects the magnetic dissipation and Poynting fluxes within the magnetosphere.
    In Appendix \ref{Sect:Appendix:Diffusion} we present an additional numerical test and a discussion on the sensitivity of our Poynting flux calculations on numerical diffusivity, which we found to be minor to the results of this study. The use of a less diffuse numerical scheme resulted in more resolved small-scale structures in plasma density and a slight increase in the auroral Poynting flux (i.e., Eq. \ref{Eq:Auroral_Poynting_flux}) by a factor of about $1.2$.

    The spherical grid, ranging from $1\,R_p$ (inner boundary) to $60\,R_p$ (outer boundary), consisted of $230$\,$\times$\,$72$\,$\times$\,$96$ grid cells in $r$, $\theta,$ and $\phi$ dimensions. The finest $r$-resolution was $0.02\,R_p$. The cell size gradually increased by a factor of $\approx 1.01$ from $1.2\,R_p$ to $12\,R_p$ and by a factor of $\approx 1.03$ from $12\,R_p$ to $70\,R_p$. The grid was equidistant in the $\phi$ direction. We transformed the grid into a Cartesian grid in post-processing. The x-axis is aligned with the relative velocity of the wind. The z-axis is parallel to the planetary magnetic moment and rotation axis. The y-axis completes the right-handed coordinate system. The longitude ($\phi$) is measured counterclockwise from the positive y-axis. The stellar wind enters through the inflow boundary with $\phi\leq 180^\circ$. More details on the coordinate system can be found in Appendix \ref{Sect:Appendix:CoordinateSystem}.


    \subsubsection{Structure of the magnetic star--planet interaction}\label{Sect:Structure_SPI}
    
    In Fig. \ref{Fig:Magnetosphere_structure} we show 2D slices of the simulated magnetosphere for a planet at several orbital distances with the basic stellar wind (B1, $B_\star=2.8$\,G) and with the stronger magnetized wind (B10, $B_\star =28$\,G). With the B1 stellar wind model, the Alfvén radius is at $0.11$\,au. Therefore, the planets with $x\le 0.1$\,au generate Alfvén wings with only one wing connecting to the star. The  plasma density is generally higher in the upstream Alfvén wing and lower in the downstream Alfvén wing. The planet at $0.2$\,au resides within a trans-Alfvénic regime, where the magnetic lobes connected with the stellar wind field show a degenerate Alfvén wing-like structure, although a bow shock without a clearly visible division between shock and magnetosheath is formed upstream of the planet. The planets with $x > 0.2$\,au show super-Alfvénic interactions, with downstream oriented magnetic lobes, a bow shock and fairly thick magnetosheath. As a result of magnetic reconnection in the magnetospheric tail, high velocity flows emerge from the downstream side.

    With the B10 wind model all considered planets reside within the sub-Alfvénic regime. Therefore, solely the size of the magnetosphere increases from $0.03$\,au to $1$\,au due to the stellar wind total pressure decay while the principle magnetospheric structure is conserved. The planet at $0.03$\,au has a vanishing magnetopause radius of about $1.2$\,$R_p$ and nearly the entire planet is permeated by reconnected (and thus open) magnetic field lines. We note that for the B10 models, we only simulated the planet up to a distance of $0.2$\,au due to numerical artifacts occurring at larger distances caused by low plasma densities.

    \section{Auroral radio emission and the RMSL}\label{Sect:Radio_RMSL_definition}
    
    For planetary auroral radio emission to occur, a mechanism to produce energetic electrons is needed. Such a mechanism is the stellar wind-magnetosphere interaction, in which magnetic reconnection accelerates electrons most commonly in the magnetotail. Those electrons gyrate around magnetic field lines as they are launched toward the magnetic poles.
    In this study we only considered stellar-wind-driven auroras. A discussion on radio emission from rotationally dominated magnetospheres can be found in \citet{Nichols2011}.
    
    \paragraph{Energetics of ECMI radio emission}
    Following \citet{Elekes2023}, we divided the conversion from stellar wind energy to auroral radio emission into separate conversion processes. The transfer efficiency of total stellar wind energy flux ($\mathcal{P}_{\rm tot}=\mathcal{P}_{\rm B}+\mathcal{P}_{\rm kin}+\mathcal{P}_{\rm th}$; Eqs. \ref{Eq:Poynting_flux_wind}-\ref{Eq:Thermal_energy_flux}) incident on the magnetospheric cross section $A_{\rm ms}$ ($P_{\rm sw}=\mathcal{P}_{\rm tot}\cdot A_{\rm ms}$) to auroral Poynting fluxes within the planetary magnetosphere ($P_a$) is characterized by the transfer function ($T$),
    \begin{equation}\label{Eq:Transfer_function_definition}
    T =  \frac{P_a}{P_{\rm sw}}\,.
    \end{equation}
    The magnetospheric cross section $A_{\rm ms}=\pi\,R_{\rm mp}^2$ is obtained from the stellar wind-magnetosphere pressure balance at the upstream magnetopause (in SI units),
    \begin{equation}\label{Eq:Magnetopause_radius}
        R_{\rm mp} = R_p B_{p}^{1/3} \left(2 \mu_0\right)^{-1/6} \left[ \rho_{\rm sw} v_{\rm 0}^2 + p_{\rm sw} + \frac{B_{\rm sw}}{2 \mu_0} \right]^{-1/6}\;,
    \end{equation}
    with the planetary equatorial magnetic field strength $B_p$.
    Subsequently, the auroral Poynting fluxes provide the electromagnetic energy to drive electron acceleration and finally the ECMI. We characterize the efficiency of electron acceleration and the ECMI mechanism using the radio efficiency $\epsilon_{\rm r}$, so that the auroral radio power $P_{\rm radio}$ can be written as
    \begin{equation}\label{Eq:Radio_power_definition}
        P_{\rm radio} = \epsilon_{r}\, P_a = \epsilon_r \, T \, P_{sw} \,.
    \end{equation}
    As a measure of electromagnetic energy transported along planetary magnetic field lines, we calculate the auroral Poynting flux $P_a$, which is formulated as the vector projection of the Poynting vector, $\vec{S}=\vec{E}\times \vec{B} / \mu_0$, onto the unperturbed planetary magnetic dipole, $\vec{S} \cdot \vec{B}_0/B_0$ \citep{Keiling2003,Elekes2023}. This formulation of the Poynting flux captures electromagnetic energy mostly associated with Alfvén waves, which play a major role in auroral particle acceleration \citep{Keiling2003,Saur2018}. We assume that all the Poynting flux $P_a$ provided to an infinitesimally thin spherical shell around the planet is dissipated within this shell by radio emission, regardless from which direction the Poynting flux is provided, so that we can calculate the total dissipated power as
    \begin{equation}\label{Eq:Auroral_Poynting_flux}
        P_a = \oint_{A}^{} |\vec{S} \cdot \frac{\vec{B}_0}{B_0}|\, \text{d} A\,,
    \end{equation}
    where $A$ is the surface area of the spherical shell with radius $R$. Our Poynting flux formulation assumes complete and irreversible dissipation due to radio emission (following \citealt{Elekes2023}; see our Appendix \ref{Sect:Appendix:Diffusion} for a discussion on the effect of numerical dissipation on the computed Poynting fluxes).
    Thus, our subsequent radio predictions will serve as upper limit estimates. The Poynting flux close to the planet only weakly depends on the radial position. At Earth and the outer planets, the electrons are usually accelerated at altitudes $\gg\,R_p$ \citep{Zarka1998,Zarka2007} and the ECMI source regions typically lie above the ionosphere at altitudes of $\gtrsim 2\,R_p$ \citep[e.g.,][]{Zarka1998,Lamy2018,Wu2026}. For simplicity, we set the radius of the integration surface to $2\,R_p$ for all models.

    Due to our formulation of the auroral Poynting flux, the integration in Eq. \ref{Eq:Auroral_Poynting_flux} is effectively done over the cross section of a northern and southern auroral oval since the integrated auroral Poynting fluxes are preferably parallel to the magnetic field close to the open-closed field line boundary (OCFB). We return to this point in Sect. \ref{Sect:Sources_radio} when discussing Fig. \ref{Fig:Radio_maps}, where we can see an impression on the distribution of the integrated Poynting fluxes.

    \paragraph{Relationship between the RMSL and the numerically derived radio emission}
    In the Solar System, the RMSL appears to reliably function only in its magnetic variant \citep{Zarka2018}. Thus, the radio power based on the RMSL is described as follows, with the stellar wind Poynting flux incident on the magnetosphere $P_{\rm B}=\mathcal{P}_{\rm B}\cdot A_{\rm ms}$ and the efficiency factor, $\eta$,
    \begin{equation}\label{Eq:RMSL}
        P_{\rm radio}^{\rm RMSL} = \eta \, P_{\rm B}\,.
    \end{equation}
    The efficiency factor has been observationally constrained to $\eta=3$\e{-3} \citep{Zarka2018}.
    According to the RMSL, the product of the efficiencies (Eq. \ref{Eq:Radio_power_definition}) should be constant ($T \epsilon_r= \text{const.}$) and thus it should follow that $T \propto \epsilon_r^{-1}$. Although the RMSL works well for the Solar System planets, it has not been validated in the more energetic and sub-Alfvénic solar wind regime (i.e., closer to the star) and therefore it is not yet clear whether the RMSL also applies to close-in exoplanets. Doubts have already been expressed as to whether the RMSL also works for close-in exoplanets and whether the estimated radio powers overestimate the true powers \citep[e.g.,][]{Nichols2016,Turnpenney2020,Cordun2025,Evensberget2025}.
    To test the behavior of the RMSL in the more energetic regime, we needed to characterize the behavior of both efficiencies, $T$ and $\epsilon_r$, for different stellar wind energy partitions. In this study we aimed to further constrain $T$ while also inferring implications for the expected behavior of $\epsilon_r$, whose physics cannot be studied by means of MHD simulations due to the microphysical processes involved in electron acceleration and the ECMI mechanism.
    
    \paragraph{Efficiency and ionospheric permeability of ECMI radio emission}
    The efficiency of the ECMI mechanism depends on how close the generated radio waves come to the resonance condition given by local plasma conditions, which depends on the gyro frequency, $f_g=eB/(2\pi m_e)$, and electron plasma frequency, $f_p=\sqrt{n_e\,e^2/m_e \epsilon_0}/(2\pi)$, both in SI units, where $m_e$ is the electron mass, $B$ is the magnetic field strength and $n_e$ the electron density (e.g., equal to the ion density). In general, ECMI emission is generated most efficiently in plasma depleted regions with $f_g>f_p$ \citep{Treumann2006}. This also holds for electron acceleration in general, while strong magnetic fields and low densities are preferable \citep{Saur2018}. The ECMI efficiency is maximal when $f_g\gg f_p$, reaching values on the order of $\approx 0.01-0.1$ \citep{Treumann2006,Lamy2011}.
    Furthermore, self-absorption of radio waves in the plasma of the source region, as well as their reflection, occurs when $f_g<f_p$ \citep{Treumann2006,Weber2017}.

    In this study we calculated the ratio $f_g/f_p$ in the expected source regions (i.e., at $2$\,$R_p$) to both characterize the ECMI efficiency and to test whether emission can escape the local plasma,
    \begin{equation}\label{Eq:Frequency_ratio}
        f_g/f_p = B \sqrt{\frac{\epsilon_0}{n_e m_e}}\;.
    \end{equation}
    We consider locations in the magnetosphere that satisfy $f_g/f_p > 1$ to be favorable for both the generation and transmission of ECMI radio emission.

    \section{The radio-magnetic scaling law in the sub-Alfvénic regime}\label{Sect:Radio_magnetic_scaling_law}
\begin{figure}
        \centering
        \includegraphics[width=1\linewidth]{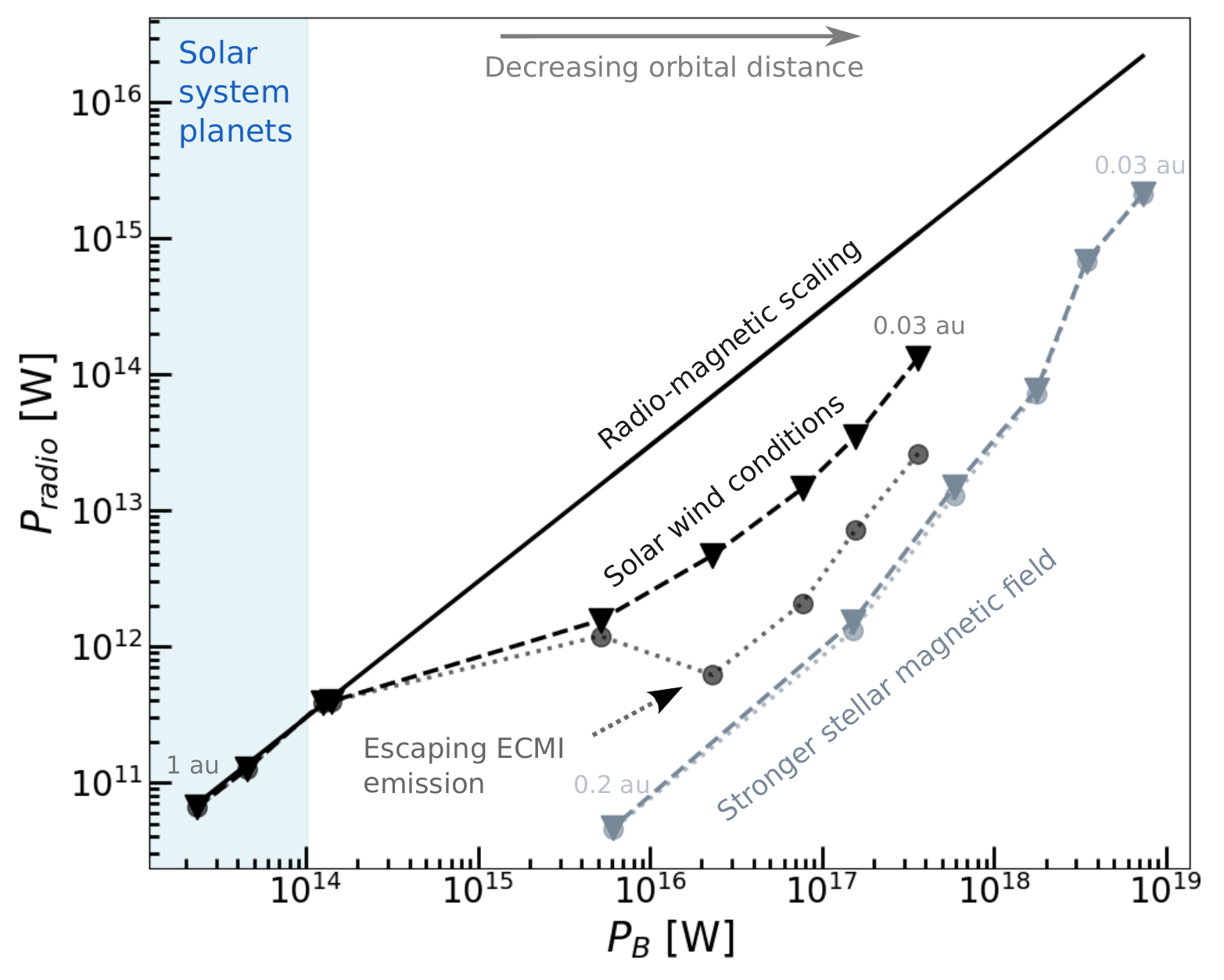}
        \caption{Predicted radio power ($P_{\rm radio}$) as a function of incident stellar wind Poynting flux times the magnetospheric cross section ($P_{\rm B}$) in watts. RMSL predictions \citep{Zarka2018} are shown as a solid line. Radio powers calculated from auroral Poynting fluxes (Eq. \ref{Eq:Auroral_Poynting_flux}) multiplied by a radio efficiency, $\epsilon_{r} = 10^{-3}$, are shown with black (B1) and gray (B10) triangles. Accompanying circles show the radio power if only escaping ECMI emission is considered. For each curve, the models range from close (rightmost points) to distant (leftmost points) planetary orbits. The blue shaded area roughly depicts the Solar System planets' regime. 
        }
        \label{Fig:Radiomagnetic}
    \end{figure}

    Following the RMSL \citep{Zarka2018}, in Fig. \ref{Fig:Radiomagnetic} we plot the predicted radio power ($P_{\rm radio}$; Eq. \ref{Eq:Radio_power_definition}) as a function of the stellar wind Poynting flux times the magnetospheric cross section $P_{\rm B}$. The radio efficiency $\epsilon_r$ is not known. We assumed that the RMSL is valid at planets within the super-Alfvénic solar wind, 
    so we could calibrate the radio efficiency in such a way that our radio predictions ($P_{\rm radio}$) and those of the RMSL ($P_{\rm radio}^{\rm RMSL}$) coincide. We assumed that the calibrated radio efficiency is the same for all planets regardless of the stellar wind regime. With a calibrated radio efficiency of $\epsilon_r=10^{-3}$, the predicted radio powers of planets in the super-Alfvénic Solar System planets regime consistently match the powers predicted by the RMSL. We used $\epsilon_r=10^{-3}$ for all further calculations in this study.
    
    In Fig. \ref{Fig:Radiomagnetic} we show radio powers with solar wind conditions (B1 model) and with the more magnetic stellar wind (B10 model) together with those predicted by the RMSL (Eq. \ref{Eq:RMSL}).
    Via calibration of the radio efficiency, the radio power in the Solar System regime (blue shaded area in Fig. \ref{Fig:Radiomagnetic}) and B1 model match the predictions of the RMSL. 
    Above an incident Poynting flux of $10^{14}$\,W (outside the blue shaded area), the B1 solar wind transitions to its sub-Alfvénic regime, from where our predicted radio powers diverge consistently from the RMSL predictions. The radio powers are reduced by one order of magnitude in the basic solar wind (B1). With the more strongly magnetized stellar wind model (B10), the radio powers drop by more than two orders of magnitude further away from the star ($>0.05$\,au). For both wind models, closer to the star the radio powers approach the RMSL predictions. In general, it can be said that in the sub-Alfvénic regime, the radio powers are significantly reduced compared to the RMSL predictions. However, it is unclear whether the assumed radio efficiency increases closer to the star due to higher reconnection rates, which was not taken into account in this study. In order for our B1 model to be consistent with the RMSL, the radio efficiency in the sub-Alfvénic regime would have to be consistently about one order of magnitude greater, $\epsilon_{r}\approx 10^{-2}$, implicitly suggesting a possible increase in the efficiency of the ECMI mechanism and the electron acceleration associated with the transition to a sub-Alfvénic interaction. For the B10 model to be consistent with the RMSL, the radio efficiency would need to be about two orders of magnitude greater above $0.07$\,au, $\epsilon_{r}\approx 10^{-1}$, which is unlikely to occur given the theoretical expectations of the ECMI and electron acceleration efficiencies \citep{Treumann2006}.
    
    We note that it is unclear whether the calibrated efficiency also applies to the B10 model, as the stronger magnetic field would influence the reconnection rate and thus the electron acceleration. However, the transfer function, which implicitly includes the energy transfer and reconnection efficiency, is similar close to the star in the B1 and B10 scenarios, and significantly reduced in the B10 model at greater orbital distances, as we show in Sect. \ref{Sect:Transfer_function}. The radio efficiency, on the other hand, is primarily determined by the magnetospheric magnetic field-to electron density ratio \citep[e.g.,][]{Treumann2006}. The planetary magnetic field remains unchanged in both models. The electron density is influenced by XUV radiation, which also remains unchanged, and plasma convection, which will be affected by the stellar wind magnetic field through reconnection processes. 
    The ECMI favorability condition (Eq. \ref{Eq:Frequency_ratio}), $f_g/f_p$ is less dependent on the electron density ($\propto n_e^{-1/2}$) compared to the magnetic field ($\propto B$), so we expect the planetary magnetic field to remain the dominant factor in radio efficiency. Hence, we expect our calibrated radio efficiency to remain within the same order of magnitude and thus to also be approximately representative for the B10 model.

    \begin{figure*}
        \centering
        \includegraphics[width=1\linewidth]{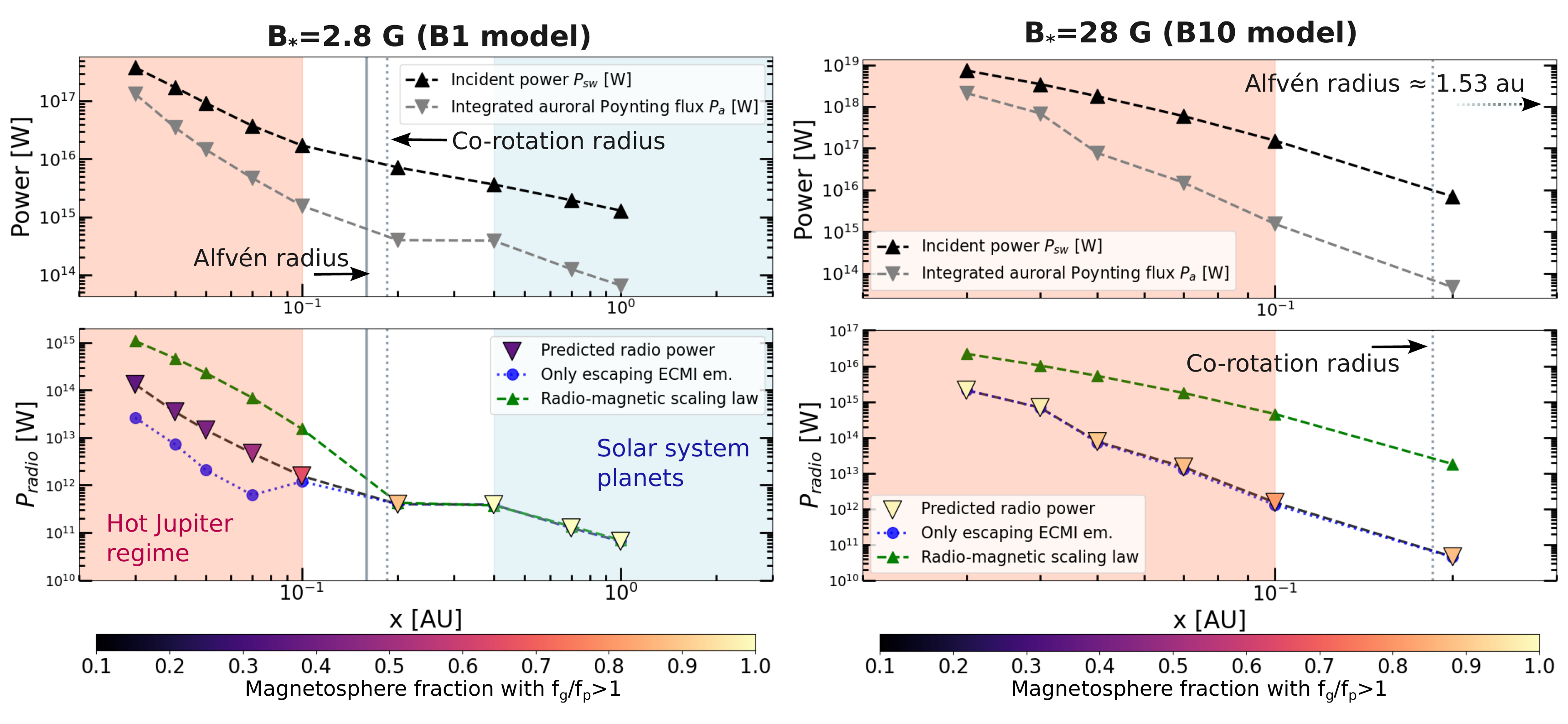}
        \caption{
        \textit{Top panels}: Integrated stellar wind energy and auroral Poynting flux at various orbital distances. \textit{Bottom panels}: Predicted radio powers according to the RMSL and based on the simulated auroral Poynting flux (colored downward triangles). Blue circles show the predicted radio power of escaping ECMI emission ($f_g>f_p$). We indicate the typical regimes of hot Jupiters (red) and Solar System planets (blue).
        }
        \label{Fig:Auroral_radio}
    \end{figure*}
    Our $P_{\rm radio}$ values implicitly assume that all the regions around the planet at $2\,R_p$ will satisfy the condition $f_g/f_p>1$ (Sect. \ref{Sect:Radio_RMSL_definition}). If we instead integrate the auroral Poynting flux (Eq. \ref{Eq:Auroral_Poynting_flux}) exclusively over regions flagged as ECMI-favorable ($f_g/f_p>1$), we get an estimate of the actual radio power escaping the magnetosphere (circles in Fig. \ref{Fig:Radiomagnetic}). Hence, we considered the ratio $f_g/f_p$ at $2\,R_p$ and flagged the associated cells accordingly. 
    
    The closer a planet is to its star, the higher the XUV-driven photoionization. The increased electron density leads to a decreasing $f_g/f_p$ ratio, associated with poorer ECMI conditions, as long as the planetary magnetic field and other sources and sinks of electrons remain unchanged. Consequently, Fig. \ref{Fig:Radiomagnetic} shows for the B1 model that radio powers further decrease by about one order of magnitude close to the star ($< 0.1$\,au, B1 model), which is consistent with the study of \citet{Weber2017}. In comparison to the RMSL predictions, our calculated radio powers are up to two orders of magnitude smaller, with this discrepancy vanishing at larger orbital distances due to weaker photoionization. The reduction in radio power drops abruptly at $0.1$\,au, even though the radio conditions generally suggest the opposite as we show later in this section. The expected radio powers of escaping emission in the B10 model are virtually unaffected by photoionization regardless of the orbital distance. This will be discussed below by taking the ECMI conditions into account.

    In Fig. \ref{Fig:Auroral_radio} we show the incident power together with the integrated auroral Poynting flux (top panels, Eq. \ref{Eq:Auroral_Poynting_flux}) used to calculate the radio powers to better illustrate the energy conversion at various orbital distances. The resulting radio powers and those predicted by the RMSL are shown as a function of orbital distance. 
    The radio powers obtained by integrating all auroral Poynting fluxes (downward triangles) are color-coded to show the volume fraction of the respective magnetospheres in which ECMI emission is likely produced and may also escape ($f_g/f_p > 1$). Darker colors indicate worse radio conditions. Accordingly, Fig. \ref{Fig:Auroral_radio} suggests that, in addition to the radio transparency of the ionosphere in close orbits (B1 model), particularly in the hot Jupiter regime, reduced ECMI efficiency can also be expected as it also depends on the $f_g/f_p$ ratio. Therefore, we expect that in reality the radio power near the star could be further reduced due to the ECMI operating at lower efficiencies. 
    In the B10 model, the expected radio powers by taking the radio conditions into account are unaffected by photoionization. On the contrary, overall radio conditions seem to be improving closer to the star, whereby the order of magnitude of $f_g/f_p$ remains the same, but the extent of the favorable regions increases. This is a consequence of the stronger stellar magnetic field and enhanced energy release during reconnection, resulting in increased reconnection rates (i.e., high velocity outflows). Stronger magnetic stresses accelerate plasma out of the magnetosphere and into the planet, where the plasma is absorbed, effectively leading to the depletion of the magnetosphere. In addition, the higher magnetic pressure of the wind results in a smaller magnetosphere and consequently smaller zones of closed field lines in which photo-ionized material can be trapped \citep{Carolan2021}. In Sect. \ref{Sect:Sources_radio} we discuss this case in more detail.

    Also visible in Fig. \ref{Fig:Auroral_radio} is the break in the trend of auroral Poynting fluxes and resulting radio power between the sub-Alfvénic and super-Alfvénic regimes (B1). The power is greatly reduced near the corotation radius where the stellar wind velocity component is parallel to the background magnetic field (see also Fig. \ref{Fig:Wind_energy}). In the B1 wind scenario, the planets at $0.2$ and $0.4$\,au exhibit approximately the same auroral Poynting flux and radio power.
    Lastly, we note that our choice of the planetary magnetic field strength has a large impact on the $f_g/f_p$ ratio and that stronger magnetic fields will enhance the local gyrofrequency. However, this assumption does not affect the general trend in the solar wind scenario that ECMI conditions are worsened by increased XUV radiation near the star.
    
    In conclusion, by placing a planet in the sub-Alfvénic stellar wind regime, we found that the predicted radio powers fall significantly below those predicted by the RMSL \citep{Zarka2018}. By considering the ECMI conditions affected by atmospheric photoionization, the predicted radio powers are further reduced if the planets are close to the star in a wind similar to the solar wind ($\ll 1$\,au). Increasing the wind magnetic field enhances the discrepancy between our radio predictions and the RMSL but at the same time reduces the impact of photoionization on the ECMI conditions due to less evaporated material being trapped in the closed magnetosphere. We used a constant radio efficiency $\epsilon_{r}=10^{-3}$ that we calibrated in the super-Alfvénic Solar System regime (B1). It is uncertain whether this efficiency is also applicable to the B10 model, but we do not expect an influence on the efficiency to overcome the power deficit of up to two orders of magnitude.
    \begin{figure*}
        \centering
        \includegraphics[width=1\linewidth]{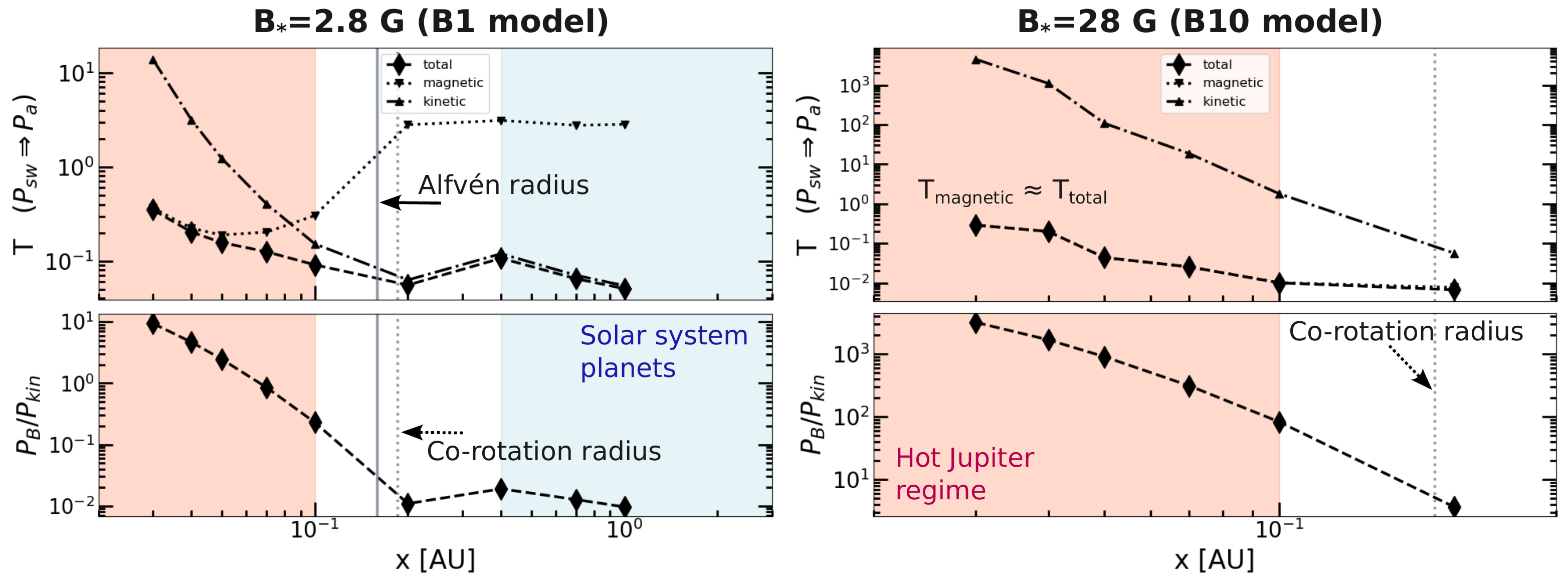}
        \caption{\textit{Top panels}: Transfer function (Eq. \ref{Eq:Transfer_function_definition}) and partial transfer functions (i.e., only considering magnetic and kinetic energy fluxes) as a function of orbital distance from the star. The partial transfer function relates to the total transfer function as $T_B^{-1} + T_k^{-1} \propto T^{-1}$. With a large partial transfer function, the corresponding wind energy flux cannot provide enough energy for the integrated auroral Poynting flux. \textit{Bottom panels}: Stellar wind Poynting flux-to-kinetic energy flux ratio.
        }
        \label{Fig:Transfer_function}
    \end{figure*}

    \section{Energetics and source regions of auroral radio emission} \label{Sect:Energetics_and_source_regions}

    In this section we discuss the energetics of the modeled magnetospheres in more detail by studying the wind-magnetosphere energy transfer (Sect. \ref{Sect:Transfer_function}). Subsequently, we take into account the spatial distribution of auroral Poynting fluxes to identify radio-favorable regions of the magnetosphere (Sect. \ref{Sect:Sources_radio}).

    \begin{figure*}
        \centering
        \includegraphics[width=1\linewidth]{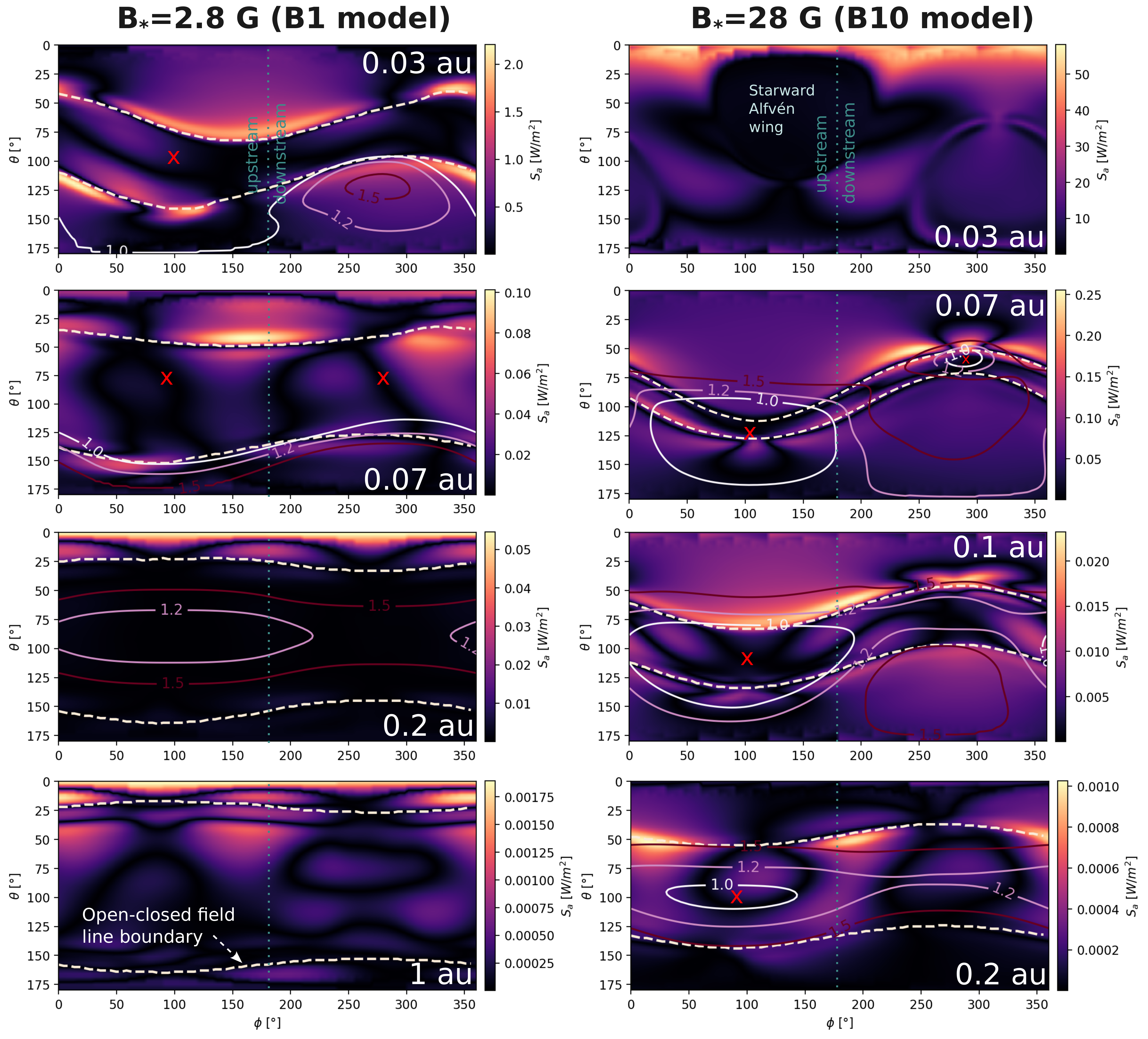}
        \caption{Mercator projection maps of auroral Poynting flux, $S_a=|\vec{S}\cdot \vec{B}_0/B_0|$ (in W/m$^2$), over a sphere with $R=2R_p$ at different orbital distances. The spatial structure of radio powers coincides with the auroral Poynting flux (Eq. \ref{Eq:Auroral_Poynting_flux}) and thus illustrates the integration across the auroral oval done in Eq. \ref{Eq:Auroral_Poynting_flux}. Dashed lines represent the OCFB. Contour lines show the $f_g/f_p$ ratio. The x and y axes show the longitude ($\phi$) and colatitude ($\theta$), respectively. The upstream side is at $\phi < 180^\circ$. Red x's mark the exterior of ECMI-favorable regions where $f_{\rm g}/f_{\rm p}<1$. A missing x indicates $f_{\rm g}/f_{\rm p}>1$ everywhere.} 
        \label{Fig:Radio_maps}
    \end{figure*}
    
    \subsection{Stellar wind-magnetosphere energy transfer}\label{Sect:Transfer_function}
    
    In Fig. \ref{Fig:Transfer_function} we show the transfer function $T$ (Eq. \ref{Eq:Transfer_function_definition}), together with the partial transfer functions, $T_B = P_a/P_{\rm B}$ and $T_k=P_a/P_{\rm kin}$, in which only the incident magnetic and kinetic powers are considered, respectively.  We note that the partial transfer functions relate to the total transfer function as $T_B^{-1} + T_k^{-1} \propto T^{-1}$, where we ignored the thermal contribution of the incident flux. 
    The bottom panels of Fig. \ref{Fig:Transfer_function} show the ratio of stellar wind magnetic to kinetic energy flux in the rest frame moving with a planet at position $x$. In the appendix we show a complementary plot of the transfer function as a function of energy partition (Fig. \ref{Fig:Appendix:T_vs_Pratio}). 
    While $T$ must always be less than one, since no energy can be produced, the partial transfer functions can. Such situations indicate that the respective incident magnetic or kinetic power cannot provide enough energy for the present auroral Poynting fluxes. Consequently, we can see from Fig. \ref{Fig:Transfer_function} that the closer a partial transfer function is to the total transfer function, the more dominant the corresponding incident energy flux is. 

    With the B1 solar wind model, the total transfer function assumes values between $0.01$ at large orbital distances and up to $0.4$ close to the star. The comparison with the magnetic-to-kinetic energy flux ratio of the stellar wind suggests a positive correlation between $T$ and the stellar wind energy partition, $P_{\rm B}/P_{\rm kin}$ (see also Fig. \ref{Fig:Appendix:T_vs_Pratio}). However, while the energy partition in the B1 model reaches $10$ at $0.03$\,au, it is two orders of magnitude larger in the B10 model, while the maximum efficient transfer function $T$ remains virtually unchanged ($T_{\rm max}\approx 0.3-0.4$ in both models). Accordingly, simply increasing the wind magnetic field strength is not sufficient to achieve higher transfer efficiencies.
    This suggests that the increased auroral Poynting flux and radio power in the B10 scenario is dominantly caused by the higher energy content of the wind, and not by a substantial increase in the energy transfer efficiency. 
    Furthermore, with the B10 stellar wind model and at distances $\geq 0.1$\,au, the transfer function drops below $10^{-2}$, one order of magnitude smaller than in the respective B1 wind scenario. This occurs at an energy partition of $P_{\rm B}\approx 10\,P_{\rm kin}$ (see Fig. \ref{Fig:Appendix:T_vs_Pratio}). At the same energy partition, the transfer function in the B10 wind scenario, $T\approx 0.08$ (at $x\approx 0.2$\,au), lies about two orders of magnitude below the respective value in the B1 solar wind scenario ($x=0.03$\,au). The transfer functions in the two wind models are only comparable at close distances to the star ($x \leq 0.04$\,au), where they reach $0.3-0.4$. At greater distances, the transfer function in the B10 case quickly decreases to values below $0.04$ (at $0.05$\,au) and below $0.01$ (at $0.1$\,au) while under solar wind conditions (B1), $T$ remains within the same order of magnitude ($0.1$-$0.2$, $x=0.04$-$0.1$\,au). Thus, our models suggest an overall less efficient wind-to-magnetosphere energy transfer with more magnetic winds.

    \subsection{Source regions and local conditions of radio emission}\label{Sect:Sources_radio}
    
    In this section we study the spatial distribution of expected auroral radio emission sources by taking into account the asymmetric distribution of Poynting fluxes and ionospheric plasma caused by photoionization.
    
    In Fig. \ref{Fig:Radio_maps} we show maps of the auroral Poynting flux per unit area (i.e., $S_a=|\vec{S}\cdot \vec{B_0}/B_0|$) on a sphere with radius $2\,R_p$ centered on the planet. 
    According to the maps, we expect radio emission to be produced preferentially at high latitudes, confined to narrow bands close to the  OCFB (dashed white lines) and mostly along open field lines. This regions are commonly referred to as auroral ovals \citep[e.g.,][]{Wu2026}. Radio powers increase with shorter distance to the star in accordance with the enhanced energy density of the wind. Intensities maximize at the flanks of the planet and on the night side. In all modeled magnetospheres, expected radio intensities peak within the northern auroral ovals, since the upstream wind magnetic field always connects to the northern hemisphere of the planet, while the southern hemisphere field connects downstream to the wind. Hence, the northern hemisphere is more open to plasma and energy injection by the stellar wind. 

    The maps show contour lines of $f_g/f_p$ ratios to indicate ECMI-favorable ($f_g/f_p>1$) and non-favorable source regions.
    In the case of strong photoionization, the favorable radio source regions are preferably found at the night side caused by lower electron densities due the absence of photoionization. In the solar wind scenario (B1), the sizes of the nightside source regions narrow the closer the planet is to the star due to the enhanced XUV flux penetrating further into the nightside hemisphere. Consequently, within our B1 model, the size of the source regions is negatively correlated with the intensity of the incident XUV flux. Furthermore, within the sub-Alfvénic solar wind, favorable source regions are exclusively found in the southern hemisphere. Plasma accumulates within the northern wing (see also Fig. \ref{Fig:Magnetosphere_structure}), while the downstream wing gets depleted by outflows partially caused by nightside reconnection.
    From about $0.1$\,au to larger distances, both auroral ovals exhibit ECMI-favorable conditions regardless of stellar irradiance. The residual influence of photoionization is only visible on the day side with increased electron densities near the magnetic equator, trapped within closed magnetic field lines (e.g., at $0.2$\,au in Fig. \ref{Fig:Radio_maps}). At greater orbital distances the entire magnetospheres exhibit favorable ECMI conditions, suggesting that atmospheric photoionization at outer planets has no significant effect on expected auroral radio emission.

    With the B10 model, the ECMI conditions are significantly improved. At $0.03$\,au, there is hardly any closed magnetosphere, so evaporated plasma can easily escape. In general, the stronger stellar magnetic field causes reconnection to be associated with higher magnetic stresses, causing strong outflows out of the radio source regions. Our results therefore suggest that planets close to the star can indeed exhibit good radio conditions despite the strong ionization of the atmosphere. The depletion of the magnetosphere, caused by the open magnetic flux and reconnection-driven outflows, dominates over plasma production via photoionization and trapping of evaporated material within closed field lines. The generally higher proportion of open field lines (evident in the lower latitudes of the OCFB) reinforces the tendency to form open channels through which plasma can escape from the magnetosphere.

    The extent of the auroral oval and thus the total auroral power is tied to the OCFB and consequently the area of open magnetic flux. In general, open flux regions enhance in size closer to the star due to the stronger wind magnetic field, with this effect being most pronounced in the strongly magnetized B10 wind. The area of open flux is consistently larger in the B10 models compared to solar wind conditions, reaching nearly 100\,\% surface coverage at $0.03$\,au. As open magnetic field lines act as channels for energy and mass transfer with the stellar wind, planets close to strongly magnetized stars generally have the largest potential for intense auroral Poynting fluxes. However, the strength of the planetary field has also an effect on the size of the open flux region, which has not been varied in our simulations.

        \section{Conclusions and summary}\label{Sect:Conclusions}
        We studied the energetics of exoplanetary auroral radio emission in sub- and super-Alfvénic stellar wind regimes at various orbital distances from the star. The main question was whether the empirical RMSL established in the Solar System can be extrapolated to close-in planets in sub-Alfvénic stellar winds. To address this question, we performed 3D MHD simulations of the stellar wind--magnetosphere interaction of a Jupiter-like planet with a $1$\,G dipolar magnetic field. We explored two wind scenarios, one similar to the solar wind and a more magnetized wind model. We calculated auroral Poynting fluxes, $P_a$ (Eq. \ref{Eq:Auroral_Poynting_flux}), the efficiency of energy transfer from the stellar wind to the auroral ovals (the transfer function, $T$; Eq. \ref{Eq:Transfer_function_definition}), and the expected radio power (Eq. \ref{Eq:Radio_power_definition}) assuming a constant radio efficiency, $\epsilon_r$. We calibrated this radio efficiency by comparing our predictions with the powers predicted by the RMSL in the super-Alfvénic solar wind, given that the RMSL is valid in the Solar System and only depends on the incident stellar wind power. With this, our calibrated radio efficiency (i.e., combined electron acceleration and ECMI efficiency) is $\epsilon_r=10^{-3}$, and we also adopted this value in the sub-Alfvénic regime due to the lack of observational constraints. In the super-Alfvénic solar wind regime, the transfer function ($T$) has to be anti-proportional to the radio efficiency ($\epsilon_r$) in order for the RMSL to hold (Sect. \ref{Sect:Radio_RMSL_definition}), which is fulfilled within our modeling. Hence, our results suggest that the combined efficiency of electron acceleration and the ECMI mechanism in the super-Alfvénic solar wind amounts to $\epsilon_r\approx10^{-3}$.

    Our results show that the auroral radio power in the sub-Alfvénic regime is consistently below the RMSL predictions (Fig. \ref{Fig:Radiomagnetic}), indicating that the extrapolation of the RMSL to close-in exoplanets may not be valid in general. This discrepancy increases with stronger stellar wind magnetic fields. We argue that, based on calculations of the ionospheric electron gyro-to-plasma frequency ratio, the radio efficiency ($\epsilon_r$) will likely not increase enough in a sub-Alfvénic wind to overcome the discrepancy between our predicted and the expected (i.e., predicted by the RMSL) radio powers. In the solar wind, our predicted radio powers are about one order of magnitude lower than the RMSL predictions, with the discrepancy increasing by up to two orders of magnitude with the more magnetized wind model.
    
    Furthermore, we studied the efficiency of transfer from stellar wind energy to auroral Poynting fluxes and found a maximum efficiency of about $0.3$ close to the star with both wind models. At greater distances, the transfer efficiency decreases to $0.05$-$0.1$ (B1) and to $0.007$-$0.04$ (B10); the lowest values are found at the greatest distances to the star. Consequently, we found a negative correlation of the transfer efficiency (i.e., the transfer function, $T$) with the strength of the stellar wind magnetic field. Hence, the transfer function ($T$) positively correlates with the stellar wind energy partition ($P_{\rm B}/P_{\rm kin}$). For the RMSL to be valid in all wind conditions, the radio efficiency must be anticorrelated with $T$, implying that the efficiency must vary by one to two orders of magnitude to achieve a constant incident energy flux-to-radio power relation (i.e., to fulfill the RMSL).
    
    We explored the spatial structure of the emission, restricting the integration of auroral Poynting fluxes to only regions of the magnetosphere where the electron gyrofrequency is higher than the electron plasma frequency. In the solar wind scenario, the expected radio power is further reduced since, although the intensities of auroral Poynting fluxes increase closer to the star, only a small fraction of Poynting fluxes are located within ECMI-favorable regions. This leads to a maximum deviation of radio powers of about two orders of magnitude from the RMSL. This effect is strongest in the sub-Alfvénic regime (i.e., near the star), where photoionization plays a major role by populating the ionosphere with electrons, which in turn prohibit the generation and escape of ECMI radio emission. 
    In the case of a more strongly magnetized stellar wind, planets very close to the star offer better radio conditions than in solar wind conditions. Due to the higher magnetic pressure, the resulting smaller magnetosphere, and the larger proportion of open planetary magnetic flux, ionized material can be carried away from the planet's auroral regions driven by magnetic reconnection more efficiently, which improves the radio conditions considerably and eliminates the mitigating effect of photoionization. However, since more strongly magnetized stars are also expected to emit higher levels of XUV, radio conditions for such stars will in the end be determined by the balance of XUV-driven electron production and the magnetic reconnection-driven electron depletion processes. Nevertheless, due to the significantly less efficient wind-to-magnetosphere energy transfer compared to the solar wind case, the expected radio powers deviate by about two orders of magnitude from the RMSL predictions even though the ECMI emission might be produced more efficiently. With regard to cool stars with stronger magnetic fields, we expect a significantly larger overestimation of radio powers by the RMSL.
    
    Our assessment of the spatial distribution of expected radio power shows that the emission maximizes within a narrow band at high latitudes, near the OCFB. Peak emission is generally expected on the night side and at the flanks of the planet. The absence of photoionization in the planet's shadow leads to a reduced electron density on the night side, considerably favoring the generation and escape of radio emission in solar wind conditions. In stronger magnetized winds, the depletion of the magnetosphere due to reconnection-driven outflows dominates over the effect of photoionization. 
    Lastly, we point out that we have not considered planet-induced radio emission from the star, which might add to the total radio power of the observed system, though emission at a different orbital phase and likely different frequencies can be expected.
    
        \begin{acknowledgements}
    This publication is part of the project VI.C.232.041 (under research programme Talent Programme Vici), which are financed by the Dutch Research Council (NWO). 
    This project has received funding from the European Research Council (ERC) under the European Union’s Horizon 2020 research and innovation programme (grant agreement No 817540, ASTROFLOW). 
    This work used the Dutch national Supercomputer Snellius with the support of the SURF Cooperative using grant no. EINF-14106. 
        \end{acknowledgements}
        
        \bibliographystyle{aa}
        \bibliography{bib/TransferFunction_paper}
        \clearpage
        
    \begin{appendix}

    \section{Structure of the stellar wind}\label{Sect:Appendix:Structure_wind}


    \begin{figure}[!htb]
        \centering
        \includegraphics[width=0.84\textwidth]{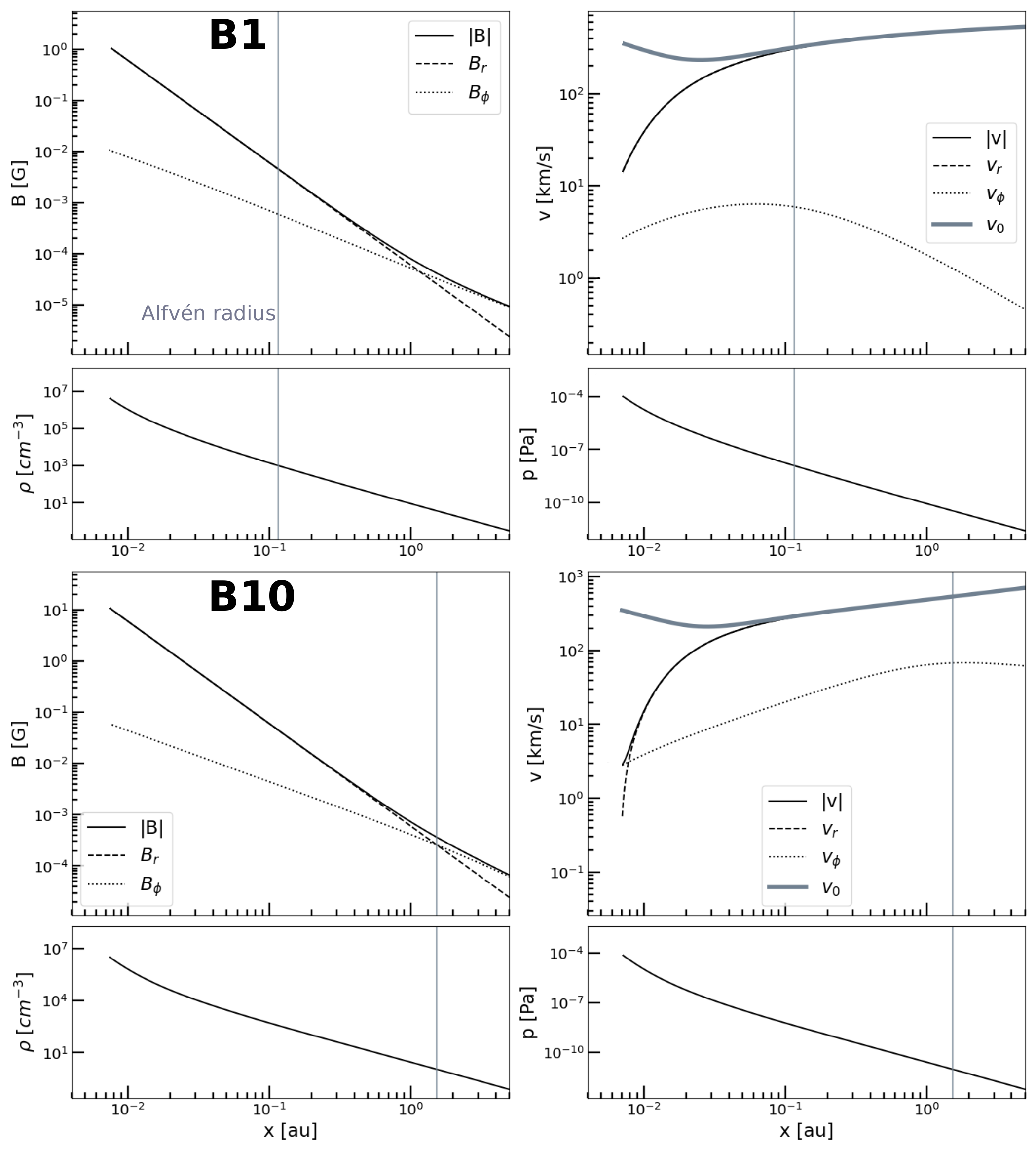}
        \caption{Stellar wind magnetic field ($B$), velocity ($v$), density ($\rho$), and pressure ($p$) as a function of distance from the star ($x$) for the solar wind (B1; \textit{top four panels}) and the more magnetic stellar wind model (B10; \textit{bottom four panels}). For the vector quantities we show the magnitude (solid lines), radial ($r$; dashed lines), and angular ($\phi$; dotted lines) components. The magnitude of the relative velocity ($v_0$) between the wind and a moving planet located at orbital distance $x$ is shown as a thick solid gray line. Vertical lines indicate the Alfvén radius.}
        \label{Fig:Stellar_wind_structure}
    \end{figure}
    
    Here we show the stellar wind properties up to $5$\,au, which are used as boundary conditions for the planetary MHD simulations. See Sect. \ref{Sect:Stellar_wind} for a description of the wind model.
    Figure \ref{Fig:Stellar_wind_structure}.1 shows the steady-state solutions of the 1D MHD simulations with the B1 (solar wind) and B10 (enhanced magnetic field) models.
    
    \clearpage

    \section{Coordinate system and geometrical definitions}\label{Sect:Appendix:CoordinateSystem}
    \begin{figure}
        \centering
        \includegraphics[width=1\linewidth]{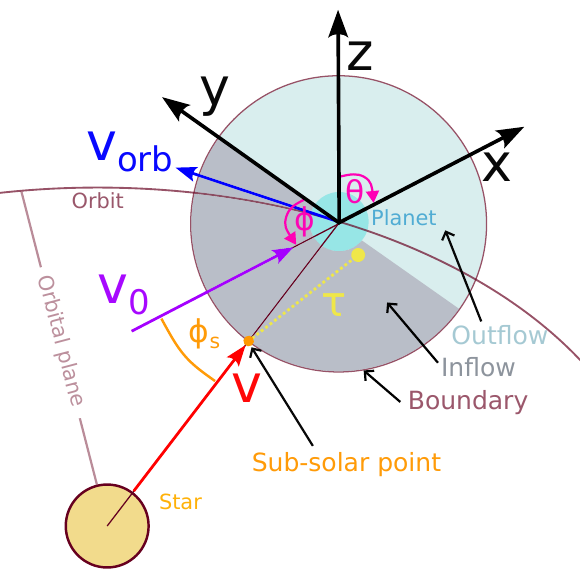}
        \caption{Sketch illustrating the coordinate systems used for computations (spherical) and the presentation of results (Cartesian).}
        \label{Fig:Appendix:CoordinateSystem}
    \end{figure}
    In Fig. \ref{Fig:Appendix:CoordinateSystem} we present a sketch demonstrating the geometrical basis of our simulation grid and computations. We assumed a Cartesian coordinate system in which the x-axis is aligned with the relative stellar wind velocity, $\vec{v}_{\rm 0}$. The relative velocity is composed of the wind velocity $\vec{v}$ and the planetary orbital velocity $v_{\rm orb}$. The z-axis is perpendicular to the orbital plane and parallel to the planetary magnetic moment. The y-axis completes the right-handed coordinate system. We solved the 3D MHD equations in spherical coordinates, where the longitude $\Phi$ is measured counterclockwise from the y-axis. The colatitude is measured from the z-axis. The inflow boundary of the spherical simulation domain is defined with $\Phi \leq 180^\circ$. The downstream hemisphere is set as zero-gradient Neumann-type outflow boundary. The optical depth $\tau$ integration (Eq. \ref{Eq:OpticalDepth}) is performed along a line starting at the substellar point toward an arbitrary cell within the computational domain. The substellar point is offset from the negative x-axis by the angular difference $\Phi_S$ between the relative velocity $\vec{v}_{\rm 0}$ and wind velocity vector $\vec{v}$. For the presentation of results (Fig. \ref{Fig:Magnetosphere_structure}), we transformed the spherical coordinates into Cartesian coordinates in post-processing.

    \section{Effect of numerical diffusion on auroral energetics}\label{Sect:Appendix:Diffusion}
    \begin{figure}
        \centering
        \includegraphics[width=1\linewidth]{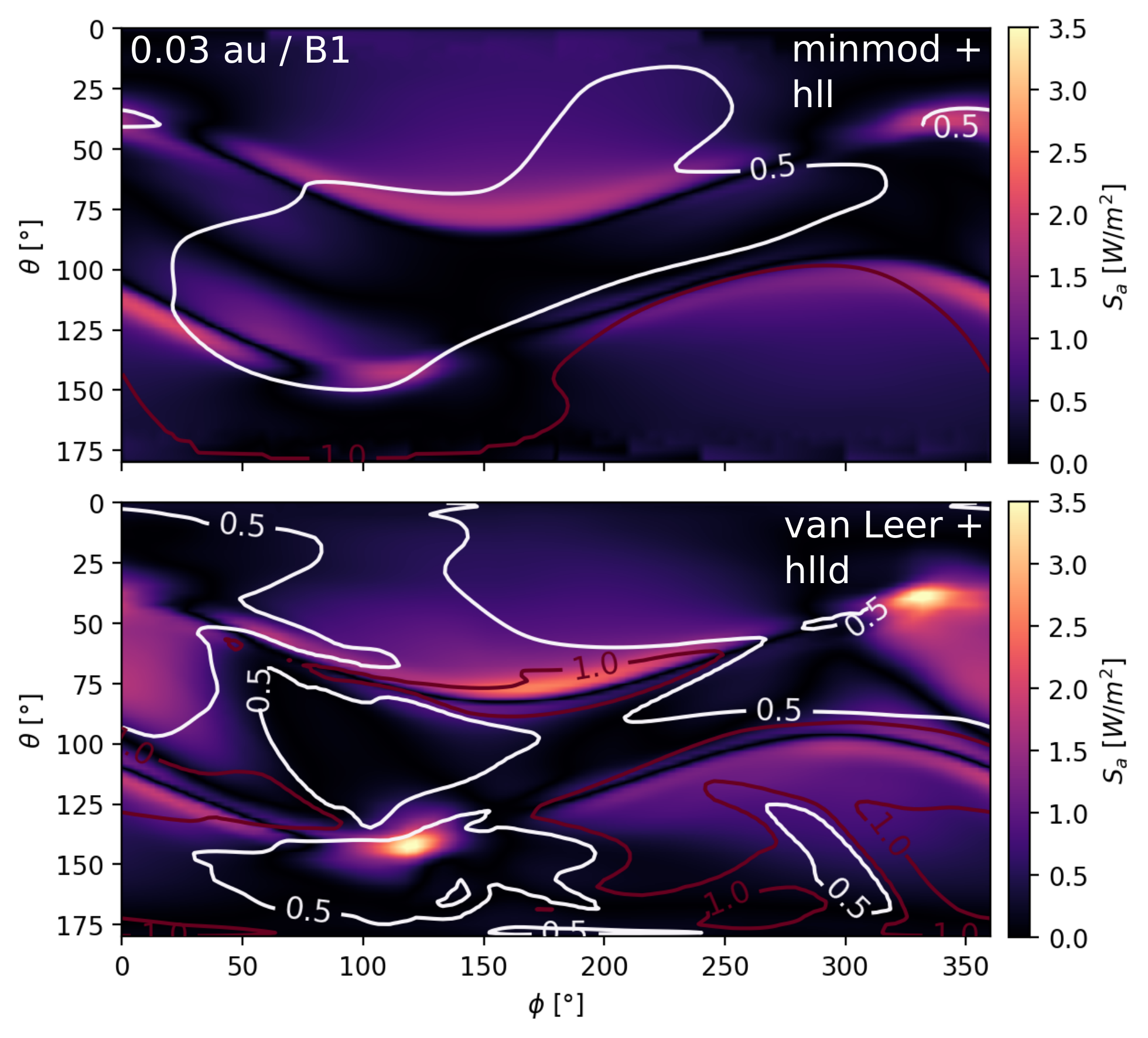}
        \caption{Poynting flux maps similar to Fig. \ref{Fig:Radio_maps} (top-left panel) for the original numerical scheme used in this study (the hll Riemann solver and the minmod flux limiter; \textit{top}) and a less diffusive scheme (hlld Riemann solver and Van Leer flux limiter; \textit{bottom}). The maps correspond to the $0.03$\,au model with the B1 solar wind model.}
        \label{Fig:Appendix:DiffusionTest}
    \end{figure}
    In this study a particularly diffusive numerical scheme has been used (i.e., the minmod flux limiter and the hll Riemann solver) in order to achieve a faster and more robust numerical solution. Hence, the dissipated magnetic energy within the magnetosphere can be affected by numerical diffusion, which may in turn affect the auroral Poynting fluxes. Here we present a brief comparison between Poynting flux computations with our  model (Sect. \ref{Sect:Planetary_MHD_model}) and with a physically equivalent model based on less diffusive numerical schemes. Instead of the minmod flux limiter, we utilized the van Leer algorithm. The Riemann problem was solved using the low-dissipation Harten-Lax-Van Leer  discontinuities (hlld)  scheme instead of the more diffusive Harten-Lax-Van Leer (hll) scheme. In Fig. \ref{Fig:Appendix:DiffusionTest} we show a comparison of the Poynting flux maps similar to Fig. \ref{Fig:Radio_maps} for the B1 solar wind model at $0.03\,$au, obtained with the more and less diffusive schemes.

    We find that the global structure of the Poynting fluxes is conserved between the two simulations. With the less dissipative scheme, however, more small-scale details of the Poynting flux and $f_g/f_p$ ratio are resolved. Furthermore, steeper gradients in Poynting flux and $f_g/f_p$ can be found, as is expected from a less diffusive numerical scheme. The most important features in the Poynting flux maps, i.e., the north-south and upstream-downstream asymmetries in Poynting flux and the auroral ovals tied to the OCFB are largely conserved. With the less diffusive scheme, narrow regions at the upstream hemisphere now show $f_g/f_p>1$ although they do not contribute significantly to the energetics.
    We performed the Poynting flux integrations (Eq. \ref{Eq:Auroral_Poynting_flux}) with the less diffuse model. The integrated auroral Poynting flux without taking the $f_g/f_p$ ratio into account amounts to $1.26\times 10^{17}$\,W and $1.47\times10^{17}$\,W for the original and less diffusive scheme, respectively. Taking the $f_g/f_p$ ratio into account the integrations yield powers of $2.5\times 10^{16}$\,W and $3.81\times 10^{16}$\,W
    with the original and less diffusive scheme, respectively. The lesser diffusive schemes thus result in slightly enhanced integrated Poynting fluxes by a factor of $\approx 1.16$ (without $f_g/f_p$ constraint) and $\approx 1.5$ (with $f_g/f_p$ constraint).

In summary, we conclude that the integrated Poynting fluxes are only slightly affected by the choice of the numerical scheme, and we consider our results robust to the numerical details. However, for a study dedicated to atmospheric escape and its detailed effects on auroral energetics, we deem the less diffuse scheme preferable due to the better resolved ionospheric plasma density.
    
    \section{Transfer function and stellar wind energy partition}\label{Sect:Appendix:TransferFunction_Pratio}
    
    \begin{figure}
        \centering
        \includegraphics[width=1\linewidth]{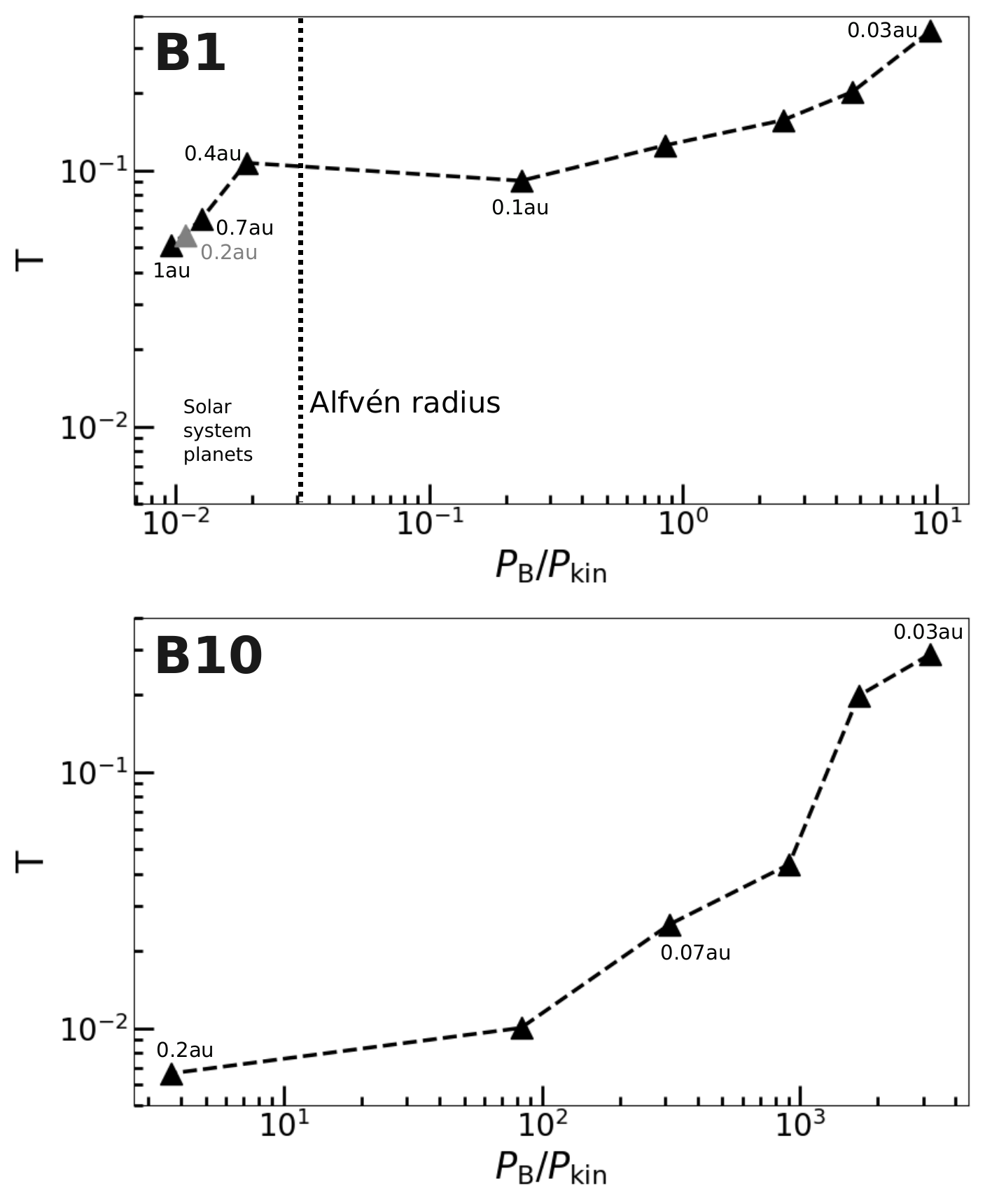}
        \caption{
        Transfer function ($T$) as a function of the stellar wind energy partition ($P_{\rm B}/P_{\rm kin}$) for the B1 (\textit{top}) and B10 (\textit{bottom}) stellar wind scenarios. Selected data points have labels indicating the orbital distance of the respective planet. In the top panel, the data point labeled 0.2 au is plotted separately due to the exceptionally low $T$ and low $P_{\rm B}/P_{\rm kin}$ caused by the proximity to the stellar wind corotation radius. The vertical dashed line indicates the Alfvén radius in the stellar frame of reference.}
        \label{Fig:Appendix:T_vs_Pratio}
    \end{figure}
    
    In this section we show the behavior of the transfer function ($T$) as a function of stellar wind energy partition ($P_{\rm B}/P_{\rm kin}$) supplementary to the transfer function plot in Fig. \ref{Fig:Transfer_function}. In Fig. \ref{Fig:Appendix:T_vs_Pratio} the different regimes of the wind can be examined and how the respective energy partition influences the transfer function. In the solar wind (B1), the transfer functions amounts to $0.1$ - $0.3$ from $0.1$\,au down to $0.03$\,au. In the super-Alfvénic wind regime, the transfer function drops monotonically to $0.05$ at $1$\,au and it is expected to drop further at larger distances. In the more magnetic wind (B10), the transfer function quickly drops from $0.3$ at $0.03$\,au down to $0.01$ at $0.1$\,au. At larger distances the transfer function drops $0.06$ and it is expected to drop further, although the curve suggests the transfer function to asymptotically approach some efficiency about $10^{-3}$.
    
    \end{appendix}
\end{document}